\documentclass[backref,twocolumn,breaklinks,colorlinks,tighten,]{aastex701}

\hypersetup{linkcolor=blue,citecolor=blue,filecolor=cyan,urlcolor=magenta}

\usepackage{enumitem}
\usepackage{mathtools}
\usepackage{graphicx}
\usepackage{amssymb}
\usepackage{amsmath}
\usepackage{ulem}
\usepackage{epstopdf}
\usepackage{color}
\usepackage{aas_macros}
\usepackage{enumitem}  
\usepackage{hyperref}
\usepackage{xcolor}

\newcommand{\chandra}{\textit{Chandra}}

\newcommand{\src}{PSR\,\,J1723\ensuremath{-}2837}

\begin{document}

\author[0000-0002-6548-5622]{Michela Negro}
\affiliation{Department of Physics and Astronomy, Louisiana State University, Baton Rouge, LA 70803, USA}
\email{michelanegro@lsu.edu}

\author[0000-0001-9826-1759]{Haocheng Zhang}
\affiliation{Astrophysics Science Division, NASA Goddard Space Flight Center,Greenbelt, MD 20771, USA}
\email{}

\author[0000-0002-7574-1298]{Niccol\`o Di Lalla}
\affiliation{Department of Physics and Kavli Institute for Particle Astrophysics and Cosmology, Stanford University, Stanford, CA 94305, USA}
\email{niccolo.dilalla@stanford.edu}

\author[0000-0002-9870-2742]{Slavko Bogdanov}
\affiliation{Columbia Astrophysics Laboratory, Columbia University, 550 West 120th Street, New York, NY 10027, USA}
\email{slavko@astro.columbia.edu}

\author[0000-0003-2714-0487]{Zorawar Wadiasingh}
\affiliation{Department of Astronomy, University of Maryland, College Park, Maryland 20742, USA}
\affiliation{Astrophysics Science Division, NASA Goddard Space Flight Center,Greenbelt, MD 20771, USA}
\affiliation{Center for Research and Exploration in Space Science and Technology, NASA/GSFC, Greenbelt, Maryland 20771, USA}
\email{}

\author[0000-0002-7465-0941]{Noel Klingler}
\affiliation{Center for Space Sciences and Technology, University of Maryland, Baltimore County, Baltimore, MD 21250, USA}
\affiliation{Astrophysics Science Division, NASA Goddard Space Flight Center,Greenbelt, MD 20771, USA}
\affiliation{Center for Research and Exploration in Space Science and Technology, NASA/GSFC, Greenbelt, Maryland 20771, USA}
\email{}

\author[0000-0002-8548-482X]{Jeremy Hare}
\affiliation{X-ray Astrophysics Laboratory, Code 662 NASA Goddard Space Flight Center, Greenbelt, MD 20771, USA}
\affiliation{University of Maryland, Baltimore County, Baltimore, MD 21250, USA}
\affiliation{The Catholic University of America, 620 Michigan Ave. NE, Washington, DC 20064, USA}
\email{}

\title{Testing Magnetic Field Configurations in Spider Pulsar \src\ with IXPE}

\begin{abstract}
We present the first X-ray polarimetry observations of a redback millisecond pulsar binary, \src, with the Imaging X-ray Polarimetry Explorer (IXPE). 
We conduct a spectro-polarimetric analysis combining IXPE data with archival Chandra, XMM-Newton, NuSTAR, and Swift observations. We explore two limiting magnetic field configurations, parallel and perpendicular to the bulk flow, and simulate their expected polarization signatures using the {\tt 3DPol} radiative transport code. To account for the rapid rotation of the polarization angle predicted by these models, we implement a phase-dependent Stokes alignment procedure that preserves the polarization degree while correcting for a phase-rotating PA. We also devise a new maximum-likelihood fitting strategy to determine the phase-dependence of the polarization angle by minimizing the polarization degree uncertainty. This technique hints that the binary may be rotating clockwise relative to the celestial north pole. 
We find no significant detection of polarization in the IXPE data, with PD~$\lesssim50\%$ at 99\% confidence level. Our results excludes the high-polarization degree scenario predicted by the perpendicular field model during the brightest orbital phase bin. Simulations show that doubling the current exposure would make the parallel configuration detectable. The new PA rotation technique is also applicable to IXPE data of many sources whose intrinsic PA variation is apriori not known but is strictly periodic.
\end{abstract}


\section{Introduction} \label{sec:intro}

A subset of non-accreting, rotation-powered millisecond pulsars in binaries, known as the black widows and redbacks (RBs), constitute one of the cleanest and well-constrained  collection of astrophysical systems for the study of pulsar winds and particle acceleration in magnetized and relativistic plasmas. In these ``spider" systems\footnote{The subclasses, the black widows \citep{1988Natur.333..237F} and redbacks \citep{2005ApJ...630.1029B}, are chiefly differentiated by their companion mass \citep{2011AIPC.1357..127R}. A third class, known as huntsmans, occupy a phase space of larger orbital periods \citep{2017ApJ...851...31S,2025ApJ...980..124S}.}, a millisecond pulsar is joined by an evolved companion in a tight and compact circular orbit with periods of typically $\lesssim$ day \citep[e.g.,][for a recent catalog]{2025arXiv250511691K}. The millisecond pulsar and its wind and companion interact in a variety of ways revealing a rich set of phenomena from the radio to the gamma-rays \citep[e.g.,][]{2022ApJ...941..199S,2025ApJ...984..146S}. This interaction offers key probes of plasma physics and the content of pulsar winds \citep{arons1993high}. The systems are well constrained in geometry and component masses from radial velocities, optical/IR spectra and pulsar timing, and seem to harbor neutron stars heavier than 1.4~$M_\odot$ \citep[e.g.,][]{2020mbhe.confE..23L}. 

Many spider binaries exhibit a non-thermal orbitally-modulated emission component in the X-ray band \citep[see][for recent compilations]{2022ApJ...941..199S,2025arXiv250511691K}, with luminosities in the range $10^{30}-10^{33}$ erg s$^{-1}$, attributed to particle acceleration in an intrabinary pulsar wind termination shock and Doppler-boosting of a bulk flow along the shock tangent \citep[e.g.,][]{2015hasa.confE..29W,2016arXiv160603518R,Wadiasingh2017,2019ApJ...879...73K,2020ApJ...904...91V,2021ApJ...917L..13K,2024ApJ...964..109S}. The existence of a shock is also indicated by orbital-phase and frequency-dependent radio eclipses of the MSP \citep[e.g.,][]{2009Sci...324.1411A}. A now well-established but surprising characteristic of all RBs is that the intrabinary shock appears to enshroud the pulsar, similar to gamma-ray binaries \citep[e.g.,][]{Dubus2010,2013A&ARv..21...64D} but with an energetically feeble companion compared to the MSP. This shift in understanding these systems has occurred in the last decade, thanks in part to X-ray characterization of numerous RBs. The X-ray orbital modulation of such systems is near maximum at pulsar inferior conjunction (when the pulsar is in front of the companion -- at phase $0.75$, if the ascending node of the pulsar is defined as phase 0). As a consequence of this shock orientation, a large fraction of the pulsar's total wind output is captured, enabling a high radiative efficiency for a given spin-down power \citep[a few percent in the $0.5-10$ keV X-ray band, cf.~][]{2015arXiv150207208R}, which may also be important for subsequently heating/irradiating the companion \citep[e.g.,][]{2023MNRAS.525.2565T,2025arXiv250310511M}. RBs are compelling TeV observatory targets \citep{Harding1990,2020ApJ...904...91V,2022icrc.confE.686W,2024ApJ...964..109S}, and possibly even are appreciable high-energy neutrino emitters \citep{2025arXiv250820952V}. 

The magnetic field structure at the shock, and where particles radiate is an open research  question.  In contrast to pulsar-wind nebulae (PWNe), whose termination shocks lie $\sim10^{15}$–$10^{17}$~cm from the pulsar, in BWs/RBs the companion orbits at $\sim10^{11}$~cm.  The much smaller standoff distance yields magnetic fields that, prior to reconnection, ought to be four–six orders of magnitude stronger than in termination shock PWNe. Confinement and the shock interface at the intrabinary shocks are also expected to be geometrically simpler than in the more messy supernova remnant-confined PWNe. This ought to lead to more ordered and highly magnetized plasmas, more efficient particle acceleration, and possibly harder particle and photon spectra.
%stricter and ``cleaner" than in the more messy supernova remnant-confined PWNe. This ought to lead to more ordered and highly magnetized plasmas, more efficient particle acceleration, and possibly harder particle and photon spectra.

There are numerous challenges simulating these systems from first principles. The plasma scales at and around the shock involved are beyond those that any particle-in-cell simulation can resolve globally with the true particle energies. The gyroradii of the highest energy electrons (e.g. those reaching TeV energies) are macroscopically large ($\sim 10^9-10^{10}$ cm), yet the plasma skin depth, particularly from the companion side, is tiny ($\ll10^3$ cm where radio eclipses occur, as implied by differential radio dispersion measure across the orbit through the shock). Yet, both must be resolved simultaneously. The system is intrinsically three-dimensional, with a curved intrabinary shock and a pulsar wind which is cylindrically symmetric \citep[e.g.,][]{2012ApJ...749....2K,Tchekhovskoy2016}. Two-dimensional global particle-in-cell models of such systems \citep{2022ApJ...933..140C,2024MNRAS.534.2551C,2025MNRAS.tmp..266C} have confirmed predictions for Doppler boosting of flow along the shock, but also yielded a large variety of plasmoids-like structures transported along the shock. These ordered plasma structures of locally-high magnetic flux may strongly influence synchrotron radiation, and perhaps polarization signals.   They are embedded within larger-scale ordered or turbulent magnetic fields. It is unclear if  plasmoid structures are sustained in reality, or what their impact is on the longer length scales probed by the synchrotron-radiating electrons in the X-ray band. For instance, where particles are accelerated and where they cool may not be co-spatial, and the pressure balance sourced from the companion may be qualitatively different than assumed in simulations \citep[e.g., a companion magnetosphere,][]{2018ApJ...869..120W}\footnote{A recent three-dimensional global particle-in-cell study of \citet{2025arXiv250811625S} considered polarization signals from BW-like systems, where the pulsar wind is assumed to be far-field and plane-parallel. In contrast, in systems where the shock envelopes the pulsar such as \src\, the pulsar wind's quasi-spherical wave fronts will influence the shock structure in a qualitatively different manner, particularly near the wings where Doppler-boosting of flux is strongest. Global particle-in-cell simulations of such a case have not yet been reported in the literature.}. Generically, higher energy particles, with their shorter energy loss timescales probe shorter length scales in the radiative region. In addition, purely geometric and relativistic effects can alter polarization signals \citep[e.g.,][]{2005MNRAS.360..869L,2021ApJ...922..260X,2023ApJ...959...81S}. X-ray polarization thus offers a powerful empirical technique to give insight on the physics involved beyond the reach of simulations.

\src\ \citep[discovered by][]{2004MNRAS.355..147F} is RB system harboring a MSP with period $P = 1.86$ ms  and $\dot{P} = 7.61\times 10^{-21}$ s s$^{-1}$, corresponding to a pulsar spin-down power of $L_{\rm SD} \sim 5\times 10^{34}$ erg s$^{-1}$, timed by \citet{crawford2013psr}. This system has an orbital period of $P_{\rm b} = 14.8$~hr and exhibits extended radio eclipses around pulsar superior conjunction. The optical companion and pulsar radial velocities yield a orbital inclination angle of $i = 30^{\circ} - 41^{\circ}$ \citep{crawford2013psr} for a range of plausible neutron star masses. Optical spectroscopy indicates the companion to have a temperature of $T_{\rm comp} \sim 6000$~K \citep{crawford2013psr}. 
A recent {\it Gaia} DR3 parallax distance of $d\approx 0.9$ kpc \citep{2023MNRAS.525.3963K} makes this system among the closest known RBs. In its X-ray properties, \src\ exhibits strong non-thermal emission and orbital modulation approaching a factor of two peak-to-trough, with higher-flux centered near pulsar inferior conjunction (see \S\ref{sec:phaseograms} below). These observations of X-ray phasing imply that the shock envelopes the pulsar. The system appears as a point-like source to \chandra~\ \citep{2014ApJ...781....6B} with no extended emission such as may arise due to a pulsar wind nebula or a bow shock. \src\ is the brightest RB system with an unabsorbed flux of $\sim 2\times 10^{-12}$~erg~s$^{-1}$~cm$^{-2}$, photon index $\Gamma_X \approx 0.9-1.1$, and negligible $N_H \sim 10^{21}$ cm$^{-2}$ \citep{2014ApJ...781....6B}.
The Imaging X-ray Polarimetry Explorer \citep[IXPE;][]{IXPEPreLaunch} observed \src~ in two sections, starting on 2024-08-20T18 and 2024-09-01T06, respectively, for a total effective exposure (livetime) of 1254335.9 s. The two observations have been merged into one OBSID (03006799).

In this paper, we present deep IXPE observations of \src\, which constitute the first X-ray polarimetric observations of a RB system. In Section~\ref{sec:theo} we describe the theoretical models and the simulations used to make predictions regarding the degree and phase dependence of the polarized emission. In Section~\ref{sec:methods} present the data analyses and the results, while in Section~\ref{sec:dicussion} we discuss the implications of the results, provide a short summary and our conclusions. 

\section{Phenomenological Polarization Models} 
\label{sec:theo}

For the orbitally-modulated X-ray emission, the observer samples a variety of Doppler factors along the shock, leading to boosting and de-boosting of flux from the wings of the shock. 
This leads to single or double-peaked X-ray light curves, with orbital modulation fractions often $>50\%$, depending on the orbital inclination of the system. Simple phenomenological models of shock emission may be constructed using a thin-shell hemispherical approximation \citep[e.g.,][]{2020ApJ...904...91V}. A zeroth-order assumption that the particle spectral index does not vary along the shock conical polar dimension leads to energy-independent light curves \citep{Wadiasingh2017}. This approximation appears to be satisfied for \src\ over the IXPE band (see \S\ref{sec:methods}).

For \src\, we consider two extreme models as a case studies in a thin-shell, hemispherical, approximation, namely, ``parallel" and ``perpendicular" field geometries.  In the parallel case, the magnetic field is parallel to the shock polar angle tangent (parallel to a prescribed bulk momentum, linearly increasing with polar angle) while it is radial in the perpendicular case. These two cases constitute the two symmetry directions in the problem; the toroidal quasi-cylindrical pulsar wind geometry, misalignment of spin and orbital axes, or a companion magnetosphere will break this symmetry, leading to significant departures of geometry and fields from axisymmetry. This would impact the polarization, possibly lowering it, but may also raise it at certain orbital phases over axisymmetric expectations \citep[e.g.,][]{2023ApJ...959...81S}. However, these cannot be accounted for while retaining generality. 

The prescribed bulk flow is a critical parameter. It is modeled as a constant acceleration in polar coordinates, i.e. a linearly increasing momentum $\beta \Gamma$ to a specified $\beta \Gamma_{\rm max}$ at $\theta_{\rm max}$ (see below). High values of $\beta \Gamma_{\rm max}$ with low $\theta_{\rm max}$ lead to widely-separated narrow peaks, while modest values of $\beta \Gamma_{\rm max}$ with large $\theta_{\rm max}$ results in a single broad peak with low pulsed fraction \citep[e.g., Figures~8--9 of][]{Wadiasingh2017}. These light curves are energy-independent if and only if the particle power-law index does not vary along the polar symmetry axis.  The linear dependence is generic (to leading order) in all pressure-confined accelerated flows, and also follows from the analytic expansions of velocity profiles in the Wilkin \citep{1996ApJ...459L..31W} and Canto \citep{1996ApJ...469..729C} thin-shell shock solutions.

In contrast to \citet{2023ApJ...959...81S}, our quantities are defined in the comoving frame and boosted to the observer frame. These geometries are implemented in the {\tt 3DPol} code \citep{2014ApJ...789...66Z,2018ApJ...862L..25Z}, which fully tracks the emission and transport of photons and their polarization from an arbitrary emitting and relativistic plasma geometry, discretized into voxels. This allows varying Lorentz and Doppler factors in time and space in the intrabinary shock with arbitrary curved geometry in our model. We do not assume pitch angle isotropy in the comoving frame, i.e. the bulk flow is assumed to describe the 3D particle distribution function in momentum completely. {\tt 3DPol} also considers light travel time effects, but these are unimportant given that the system size $10^{11} {\rm \, cm}/c \ll P_{\rm b}$.

\begin{figure*}[]
    \centering
    \includegraphics[height=7cm]{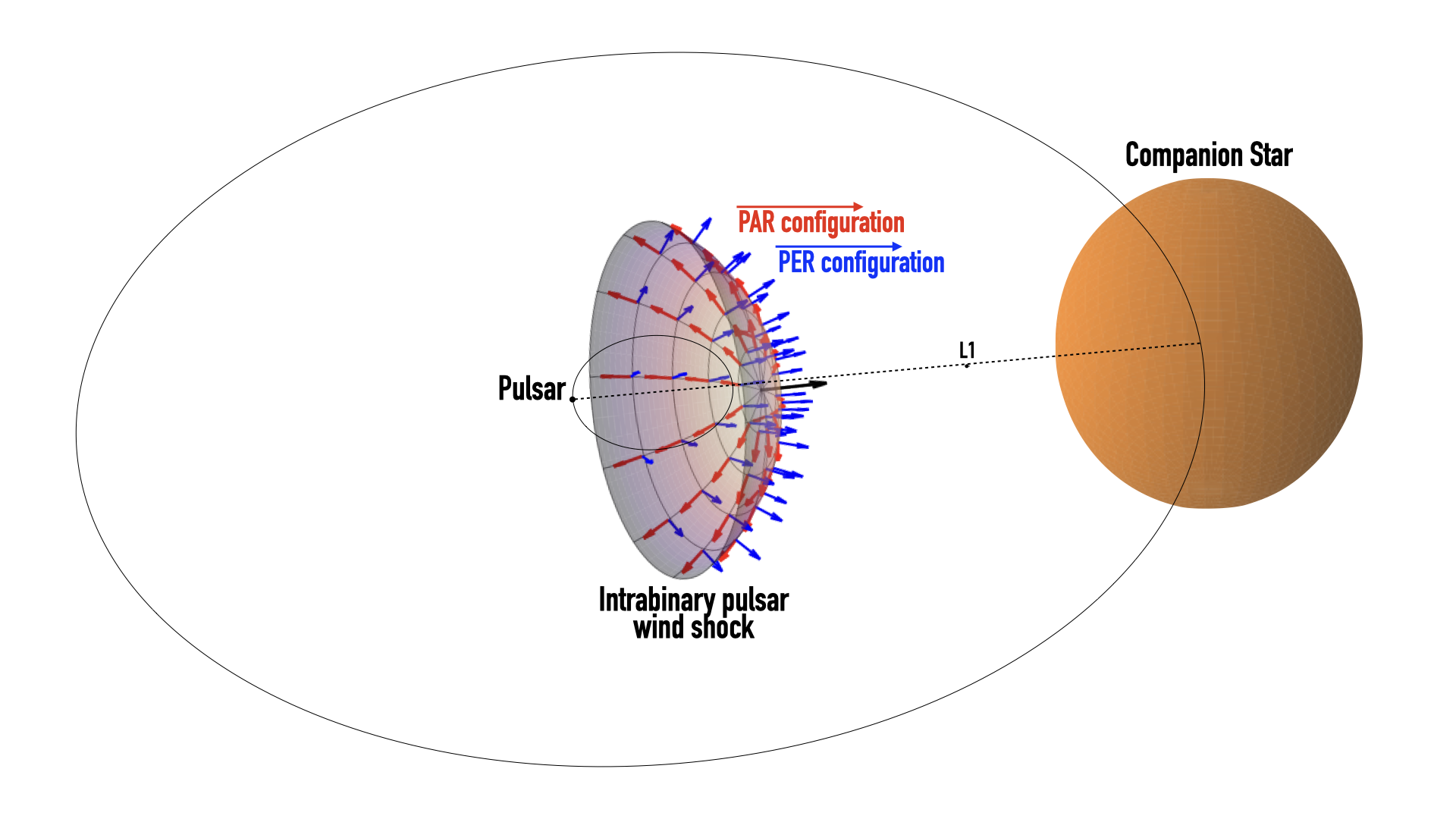}
    \includegraphics[height=8cm]{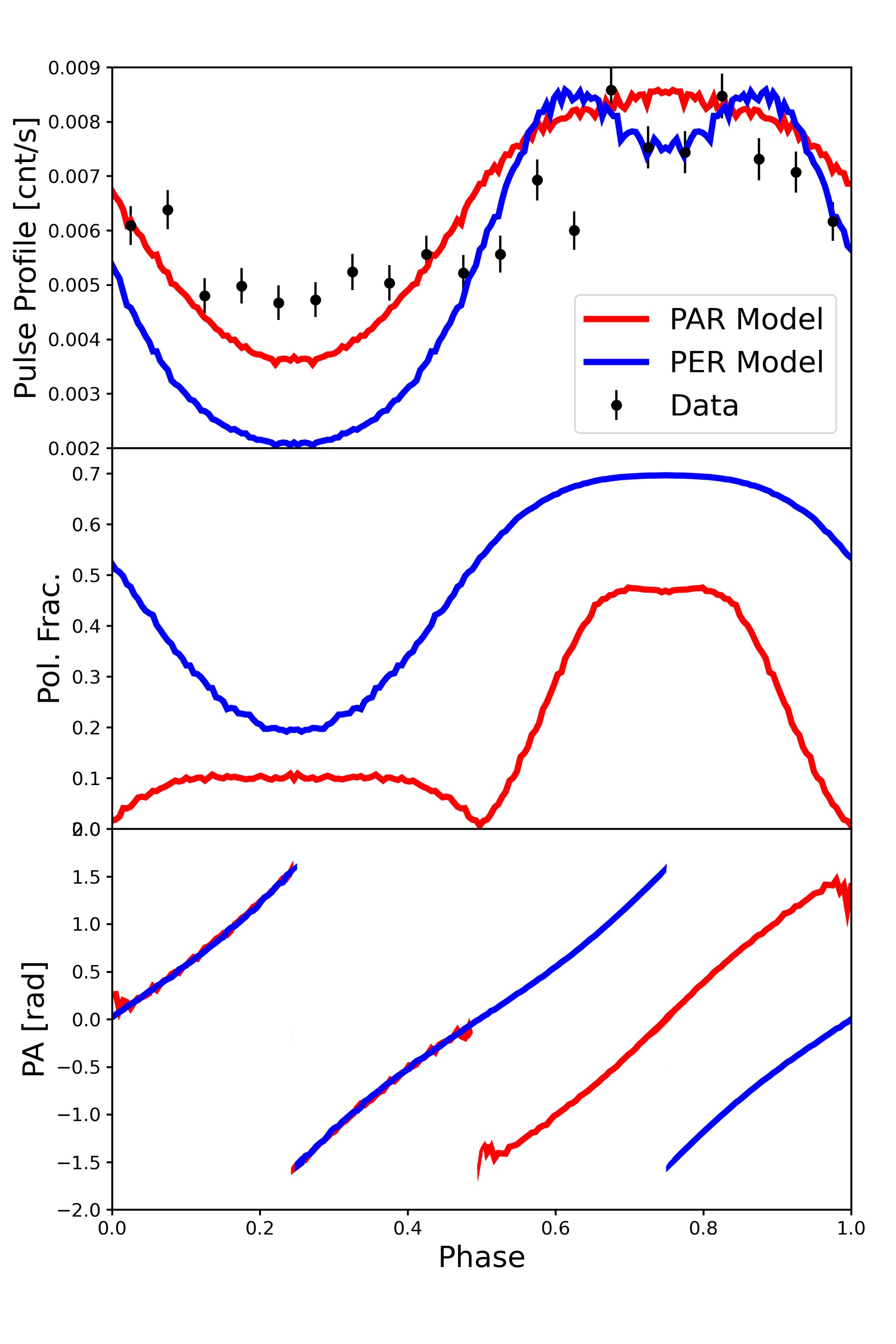}
    \caption{\textit{Left:} Schematic illustration of an intrabinary pulsar wind shock in a compact binary system as seen by an observer at inclination $i=40^\circ$ with respect to the circular orbital plane. The pulsar launches a relativistic wind that is confined by the presence of a companion star, forming a shock surface (gray-ish cap). Red arrows indicate the PAR magnetic field configuration, while blue arrows mark the PER configuration. \textit{Right:} Comparison between two polarization models for PSR J1723$-$2837 using {\tt 3DPol}: the parallel (PAR) model shown in red, and the perpendicular (PER) model shown in blue. The top panel displays the pulse profile in arbitrary units, where both models are compared against real IXPE data points. The center panel shows the polarization fraction predicted by each model, with the PER model exhibiting a significantly higher modulation amplitude than the PAR model. The bottom panel presents the predicted polarization angle (PA) as a function of phase. The discontinuous jump in the polarization position angle occurs when the net polarization fraction approaches zero, at which point the PA becomes formally undefined and may change abruptly by 90°. In the PAR configuration, this condition is naturally reached at orbital phases 0 and 0.5 due to nearly complete cancellation of the polarized emission from different regions of the shocked flow. In the PER configuration, a similar behavior is in principle possible, but it is significantly more difficult to realize. In this case, Doppler boosting is intrinsically coupled to the magnetic field geometry, and the regions where polarization cancellation can occur are confined to the wings of the shock, where the emission is strongly de-boosted. As a result, the total polarized flux rarely approaches zero, and the PA therefore evolves smoothly with orbital phase for a wide range of parameter choices.}
    \label{fig:models}
\end{figure*}

For both the parallel and perpendicular case, we survey parameters such as the bulk Lorentz factor scaling ($\beta \Gamma_{\rm max}$ between 1.1 and 2) and hemispherical maximum shock angle $\theta_{\rm max} \in \{30^\circ, 40^\circ\} $ until a rough match with the Stokes I, as in Figure~\ref{fig:models}, is obtained. For \src\ in {\tt 3DPol}, we set a box $200^3$ resolution, shock thickness of $10^{10}$ cm in the box of side $5\times 10^{11}$~cm for orbital separation $a\approx 2.8 \times 10^{11}$~cm. The shock radius, centered at the pulsar, is $0.4a$\footnote{The radius is chosen to to match the orbital phase-averaged hard X-ray spectral energy distribution of \src. In the transport model of \cite{2020ApJ...904...91V}, the size scale influence the normalization, i.e. number of emitting particles and their residence time. This is connected to pulsar parameters such the magnetization at the shock, and Goldriech--Julian particle rate and multiplicity. In the hemispherical approximation adopted here, the radius does not materially impact the light curve shape.}. The observer inclination angle is set at $30^\circ$. The simulation geometry is defined as follows: the binary orbital axis is initially oriented along the x-axis, and the observer’s line of sight lies in the x–z plane. We define the reference polarization angle (PA = 0) such that it corresponds to polarization aligned with the projection of the x-axis onto the plane of the sky, which is perpendicular to the line of sight. The magnetic field in the comoving frame is set to $1.1$~G and particle spectral index set to $-2$. The value of the magnetic field, the shock radius and pulsar parameters (i.e. tuning downwards from the maximum magnetic field allowed for toroidal pulsar wind at a set distance), sets the normalization of the SED \citep[see][]{2020ApJ...904...91V} and does not materially impact the light curve shape. These parameters are chosen roughly to be comparable to the previous study by \cite{2020ApJ...904...91V}, which considered the absolute normalization of the spectral energy distribution of \src\ in a Boltzmann transport model. Parameter degeneracies have not yet fully explored, and indeed a polarization detection may lift some degeneracies. MeV detections or limits are also necessary to constrain the particle acceleration at the shock. However, the main unknown is the structure of the magnetic field along the shock and that cannot be explored parametrically without a polarization detection.

One key feature worth noting is that the ``parallel'' case generally results in a lower polarization degree than the perpendicular case in Figure~\ref{fig:models}. This, in essence, arises from the fact that Doppler-boosted synchrotron sampled by the observer in the wings of the shock is less linearly polarized, and more elliptically polarized, than other viewing directions. Such a depolarization is also apparent in some of panels of \citet{2025arXiv250811625S}. Differences in the light curve of Stokes I in Figure~\ref{fig:models} arise from the contribution of different pitch angles the observer samples convolved with the Doppler boosting. For instance, for the PAR case, near head-on shock tangent regions that an observer samples are close to maximum Doppler boosting also have lower flux due to a small pitch angle. This is not true for the PER case, which results in more prominent double peaks at these double phases.

\section{Data Analysis and Results} 
\label{sec:methods}

\subsection{Phaseograms}
\label{sec:phaseograms}

To study the phaseogram of the target, we make use of \src\ observations by the \textit{XMM-Newton} European Photon Imaging Camera \citep[EPIC,][]{2001A&A...365L..27T} from 2011 March and the \chandra~ Advanced CCD Imaging Spectrometer (ACIS) from 2012 July originally presented in \citet{2014ApJ...781....6B}, as well as an exposure from the Nuclear Spectroscopic Telescope Array Mission \citep[NuSTAR,][]{2013ApJ...770..103H} acquired in 2015 October. Furthermore, we utilized many archival observations by the Neil Gehrels Swift Observatory \citep{2004ApJ...611.1005G}. Details of the data selection and reduction for each instrument are provided in Appendix \ref{app:data}, where we also provide Table~\ref{tab:xraylog}, summarizing the X-ray observations of PSR J1723$-$2837 used in this work.

For each data set, we extracted the phaseogram and the spectra which are reported and discussed in the following section. The time stamps of all events were barycentered using the source coordinates from \citet{crawford2013psr} obtained from radio timing ($\alpha = $ 17:23:23.1856, $\delta = -$28:37:57.17) and assuming the JPL DE405 Solar system ephemeris.

\begin{figure}
\centering
\includegraphics[trim={0.2cm 6.0cm 0.0cm 0},clip,width=0.47\textwidth]{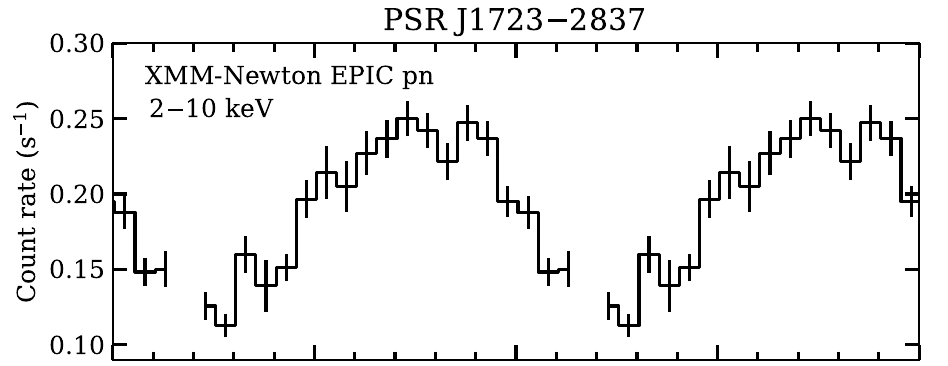}\\
\includegraphics[trim={0.2cm 6.0cm 0.0cm 0.7cm},clip,width=0.47\textwidth]{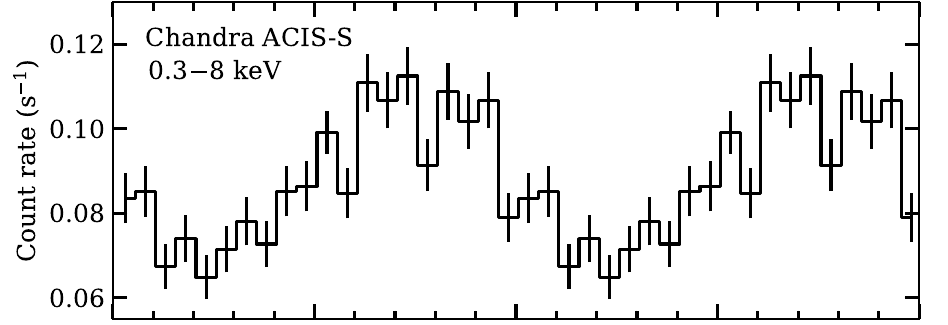}\\
\includegraphics[trim={0.2cm 6.0cm 0.0cm 0.7cm},clip,width=0.47\textwidth]{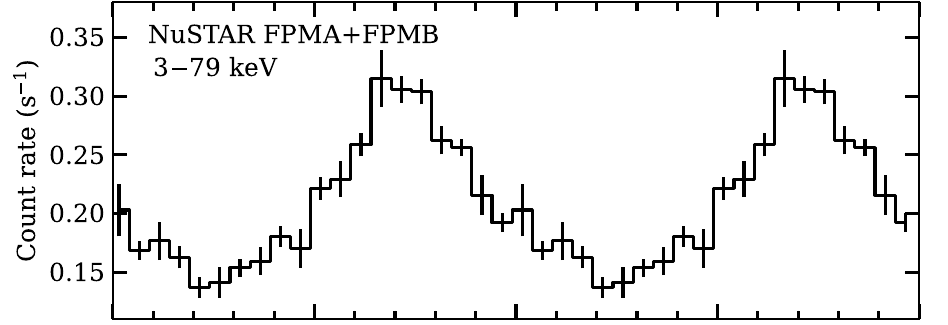}\\
\includegraphics[trim={0.2cm 5.0cm 0.0cm 0.7cm},clip,width=0.47\textwidth]{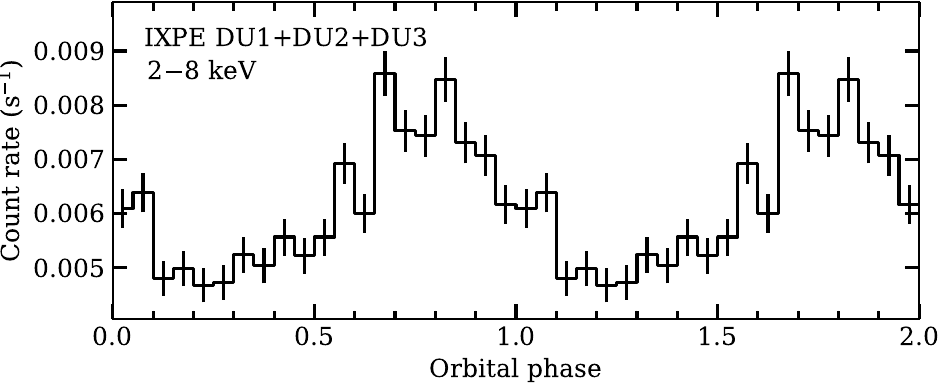}
\caption{Background subtracted X-ray light curves of PSR J1723$-$2837 obtained with XMM-Newton, Chandra, NuSTAR, and IXPE (from top to bottom, respectively) folded at the binary orbital ephemeris from \citet{crawford2013psr}. The gap in the XMM-Newton light curve is due to incomplete coverage of the orbit. Orbital phase zero corresponds to the time of the ascending node of the pulsar.  Two cycles are shown for clarity.}
    \label{fig:lcurves}
\end{figure}

Figure~\ref{fig:lcurves} shows the background subtracted light curves of \src\ from XMM-Newton, Chandra, NuSTAR, and IXPE, folded at the binary orbital ephemeris from \citet{crawford2013psr} and with 20 phase bins per cycle. The IXPE light curve exhibits a similar variability pattern as the archival X-ray observations, with all light curves showing a factor of $\sim$2 luminosity enhancement around inferior conjunction (when the pulsar is between the companion and observer).

The observed phase-dependent modulation provides an important constraint for our simulations. Any physically plausible emission model must reproduce the shape and amplitude of the observed light curve. As such, the light curve acts as a critical benchmark for validating simulated scenarios (see the top panel in fig~\ref{fig:models}). In the following section, we use the phraseogram to verify that our simulated emission is in phase with the observed variability and that it captures the overall structure of the measured light curve with reasonable accuracy.

Additionally, using the IXPE-phaseogram and the expectation of polarization signal from the models, we defined an \textit{a priori} optimal phase binning to maximize statistical significance of a polarization signal. We define two macro-phase bins 0.05--0.45 and 0.55--0.95.

\begin{figure}
\centering
\includegraphics[width=0.47\textwidth]{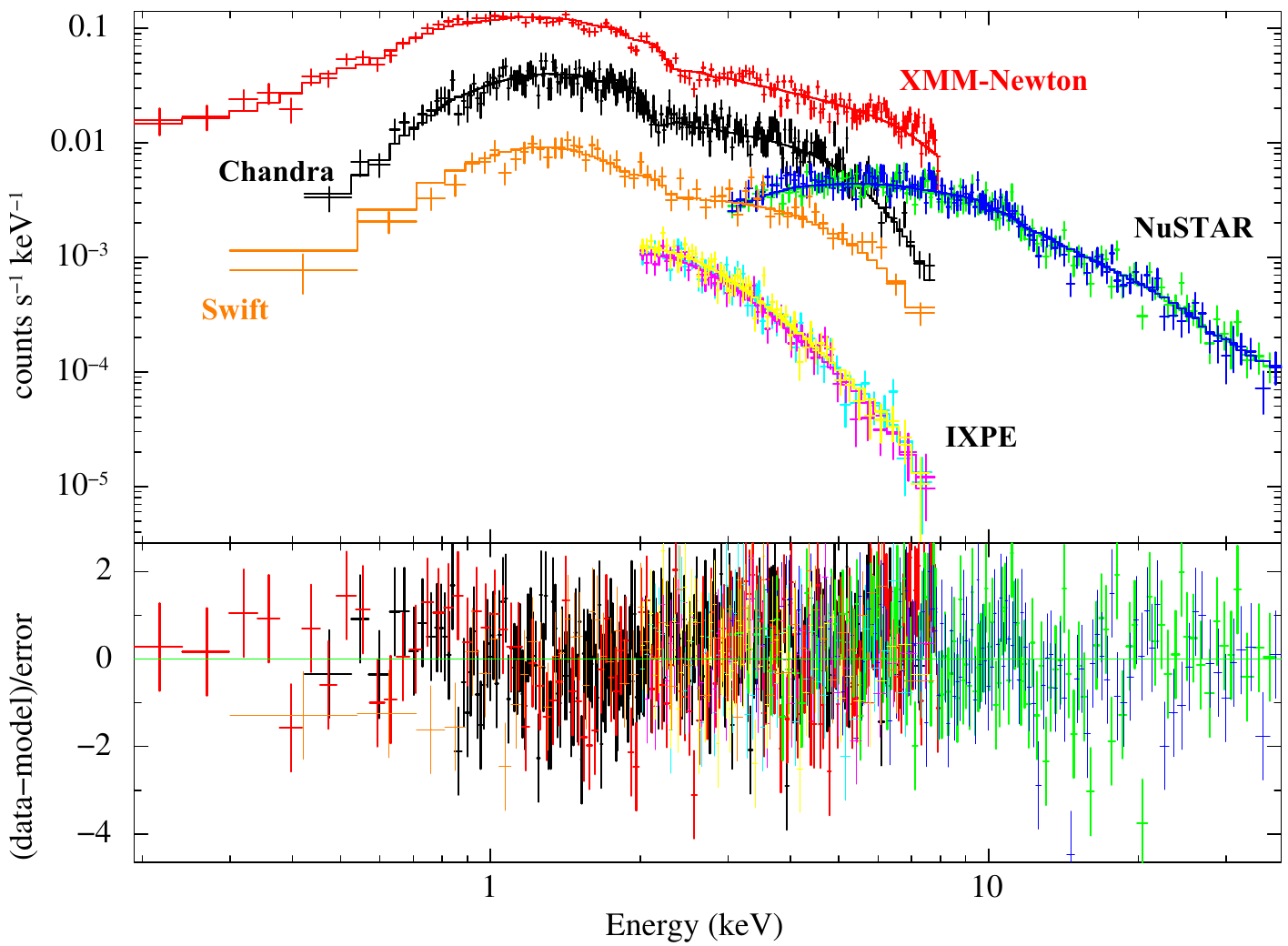}
\caption{Joint fit of the IXPE DU1--DU3 (cyan, magenta, and yellow), NuSTAR FPMA+FPMB (blue and green), Chandra ACIS-S (black), Swift XRT (orange), and XMM-Newton (red) with an absorbed powerlaw with a tied power-law index but independent normalizations. See Table~\ref{tab:fit} for best fit parameters. The bottom panel shows the best fit residuals expressed in terms of standard deviations ($\sigma$).}
    \label{fig:spectra}
\end{figure}

\subsection{Spectroscopy}
\label{sec:spectroscopy}

The phase-integrated spectra of \src\ from XMM-Newton EPIC pn, Chandra ACIS-S, NuSTAR FPMA+FPMB, Swift XRT, and IXPE DU1-DU3 were fitted jointly in XSPEC version 12.15.0. 
The fit was conducted by assuming an simple absorbed power-law model, where the hydrogen column density $N_{\rm H}$ and spectral photon index $\Gamma$ parameters are the same between all data sets. We used the \texttt{tbabs} interstellar absorption model assuming elemental abundances from \citet{2000ApJ...542..914W}.  We allow the normalization for each instrument to be an independent free parameter to allow for expected differences in the cross-calibration between X-ray instruments as well as different orbital coverage and intrinsic changes in the source luminosity between observations. 

Taking $N_{\rm H}$ to be a free parameter we find $\Gamma = 1.18 \pm 0.02$,
in agreement with previous studies \citep{2014ApJ...781....6B}. Figure~ \ref{fig:spectra} shows the results of the fit and Table \ref{tab:fit} reports the best-fit values of the assumed spectral model parameters.

\begin{deluxetable}{lC}
\tablewidth{0pt}
\centering    
\tablecaption{Summary of spectroscopic analysis for \src}
\label{tab:fit}
\tablehead{\colhead{Parameter} & \colhead{Value (XSPEC)}
          }   
\startdata
\hline
    \multicolumn{2}{c}{\textit{IXPE}} \\                
\hline
$N_H$ (10$^{21}$\,cm$^{-2}$) &  2.85 ({\rm fixed})\\
$\Gamma$                     & 1.30\pm0.06  \\
$PL ~{\rm ampl.}$ ($10^{-4}$cm$^{-2}$\,s$^{-1}$\,keV$^{-1}$) & 1.97\pm0.14\\
%$F_X$\tablenotemark{a} ($10^{-12}$\,erg\,cm$^{-2}$\,s$^{-1}$) & & \\
$\chi^2/{\rm d.o.f.}$  & 138.0/172  \\
\hline
    \multicolumn{2}{c}{\textit{IXPE+CXO+NuSTAR+Swift+XMM}} \\
\hline
$N_H$ (10$^{21}$\,cm$^{-2}$) & 2.85\pm0.10 \\
$\Gamma$                     & 1.18\pm0.02 \\
$PL ~{\rm ampl.}$ ($10^{-4}$cm$^{-2}$\,s$^{-1}$\,keV$^{-1}$) & 1.73\pm0.05 \\
%$F_X$\tablenotemark{a} ($10^{-12}$erg\,cm$^{-2}$\,s$^{-1}$) & & \\
% $PD(u.l.99\%)$& & 17.48\\
$\chi^2/{\rm d.o.f.}$  & 806.1/816 \\
\enddata
\tablecomments{ All uncertainties quoted are at a 1$\sigma$ confidence level. A cross-calibration constant has been left free to vary for all spectra, except for DU1 (whose value has been set to 1): 0.88 (IXPE DU2),  0.93 (IXPE DU3), 2.1 (\textit{Swift-XRT}), 1.52 (\chandra~), 1.98 (NuSTAR 1), 2.11 (NuSTAR 2), 2.25 (XMM).} 
%\tablenotetext{a}{Unabsorbed flux in units of $10^{-13}$\,erg\,cm$^{-2}$\,s$^{-1}$ in the \textcolor{red}{?? keV range}.}
\end{deluxetable}

\subsection{Phase-dependent Polarization analysis}

According to our models, we expect the polarization degree and angle to be a function of the orbital phase. Thefore, phase-folding and phase-binning of the IXPE data is needed to study the polarization signature. Due to the limited number of counts accumulated by IXPE during its observation, we minimized the number of phase bins in which we study the polarization, by defining two macro phase bins: 0.05--0.45, 0.55--0.95.
According to our simulations based on phenomological models \S\ref{sec:theo}, we expect big variations of the polarization angle within the large phase bins, while polarization degrees to remain relatively constant in the same phase bin. For this reason we developed a custom phase-Stokes-alignment procedure to account for such rotation. This Stokes alignment is performed on an event-by-event basis, rotating each photon by an angle computed from the Stokes parameters predicted by the physical model at a given phase. The rotation angle is thus defined as:
\begin{equation}
    \phi_{\rm rot} (\varphi) = \arctan_2 \left(\frac{U_{\rm theo}(\varphi)}{Q_{\rm theo}(\varphi)}\right)
\end{equation}
where $\varphi$ is the phase at which the stokes parameters are evaluated. Effectively, we are rotating the (Q, U) point to lie along the axis $Q=0$, keeping the same global PD, but aligning the global PA to 0, assuming as reference the expected direction and a given phase provided by our model. Thus, this procedure assumes prior knowledge of PA variation and aims to enhance signal-to-noise to retrieve PD. The aligned Stokes parameters are then:
\begin{equation}
        Q' = Q {\cos}(-2\phi_{\rm rot}) - U {\sin}(-2\phi_{\rm rot})\\
\end{equation}
\begin{equation}
    U' = U {\sin}(-2\phi_{\rm rot}) + Q {\cos}(-2\phi_{\rm rot})
\end{equation}
More details on this phase-alignment procedure are provided in appendix \ref{app:align1}. Additionally, in appendix \ref{app:align2}, we illustrate how we use this alignment procedure recursively, in a maximum likelihood approach, with the goal of inferring the intrinsic PA rotation model that minimizes the uncertainty on the estimation of the PD.

To validate the phase rotation procedure we simulate IXPE observations using the PAR and PER theoretical models described in Sec.~\ref{sec:theo}, for different multiples of the exposure of the real IXPE observation: $\times$1 (to make sure that we do match what we observe),  $\times$2 (to understand if doubling the exposure we could detect the PD predicted by the PAR model), and $\times$10 the exposure (to clearly test that our procedure works well in an ideal case of large photon statistics). The first two cases are presented here, while the latter is used primarily for illustrative purposes--- as it allows for finer phase binning and better highlights the effects of rotation--- is reported in appendix~\ref{app:align1}.

The simulation has been carried out using the \texttt{ixpeobssim} simulation suit and is based on PAR and PER theoretical models described in Sec.~\ref{sec:theo}. We used the pulsar’s orbital ephemeris available in the literature ($\text{epoch} = 55667 \text{MJD}, \nu_0 = 1.9 \times 10^{-5}$~Hz), which are the same used for the data analysis (see Sec.~\ref{sec:methods}), and the observed power-law spectrum by \chandra~ with photon index of $\Gamma=-1.13$ and matching an energy flux of $1.29\times10^{-12}$ erg/cm$^2$/s measured in the Chandra energy band (0.3-10 keV).

\begin{figure*}[h]
    \centering
    \includegraphics[width=0.45\linewidth]{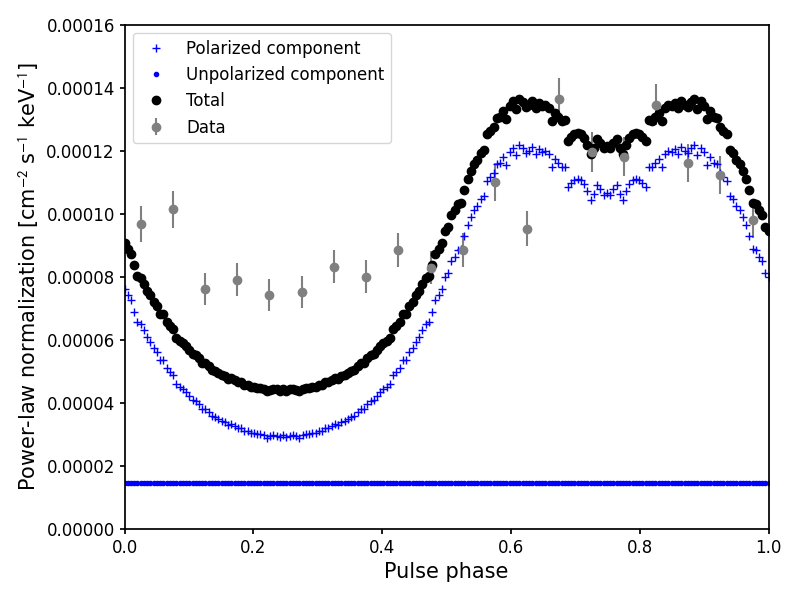}~~~~
    \includegraphics[width=0.45\linewidth]{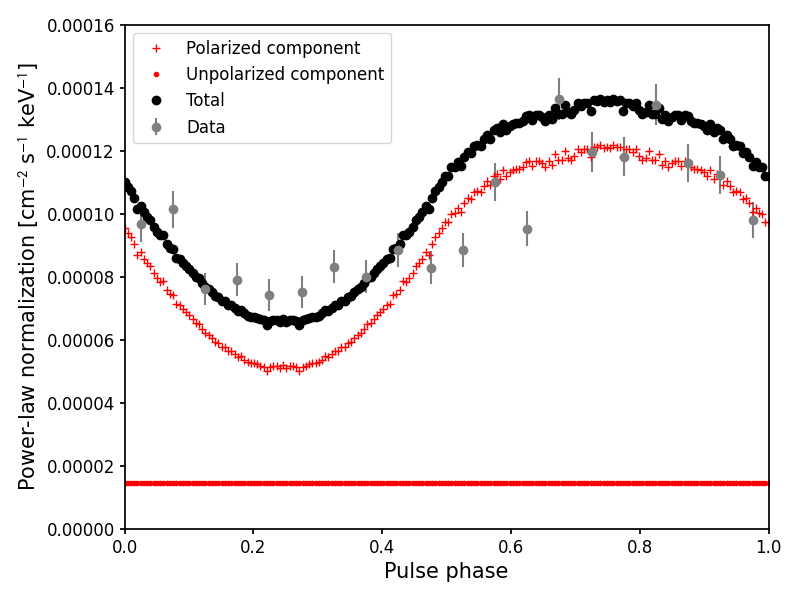}
    \includegraphics[width=0.45\linewidth, height=6cm]{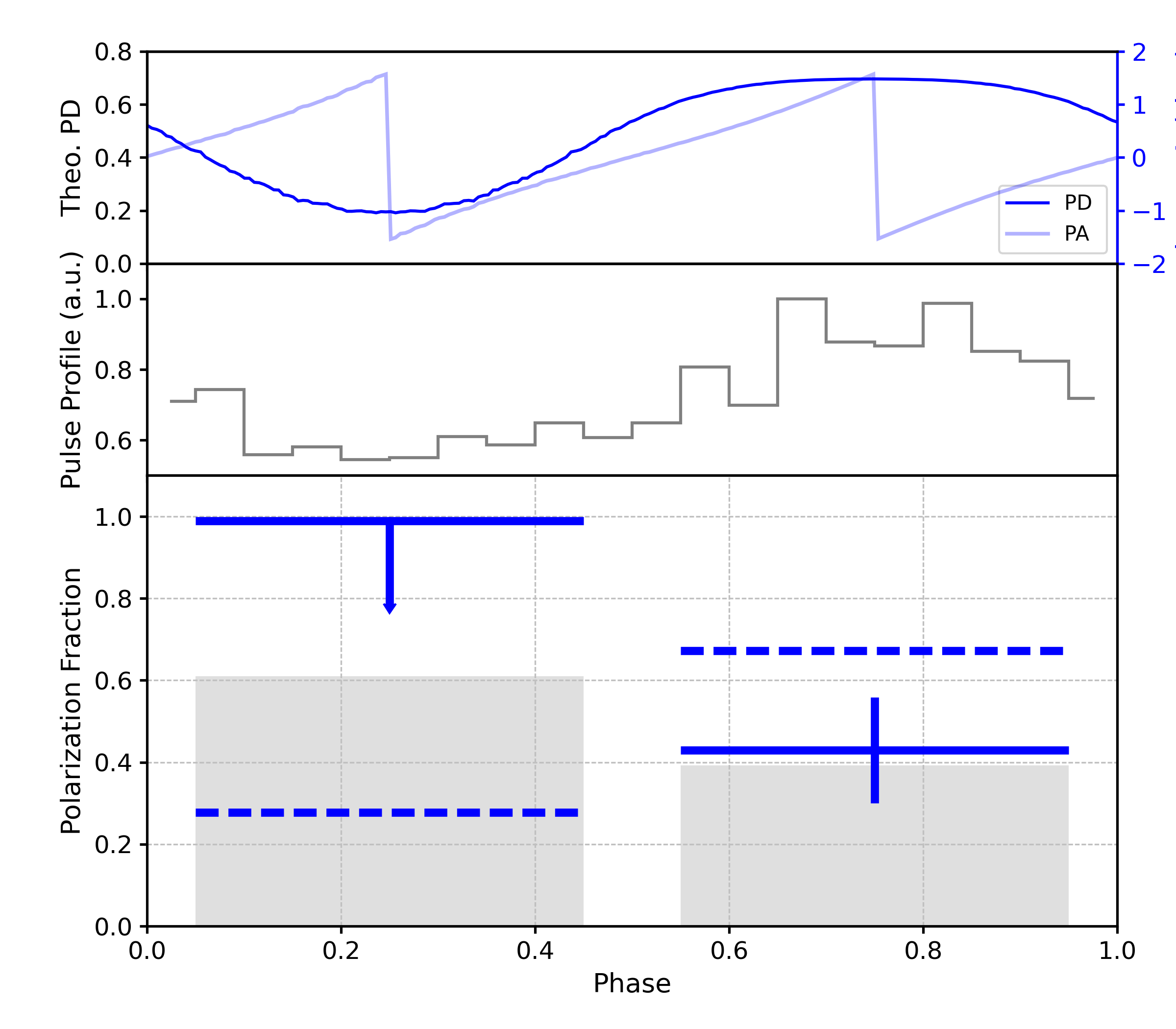}~~~~
    \includegraphics[width=0.45\linewidth, height=6cm]{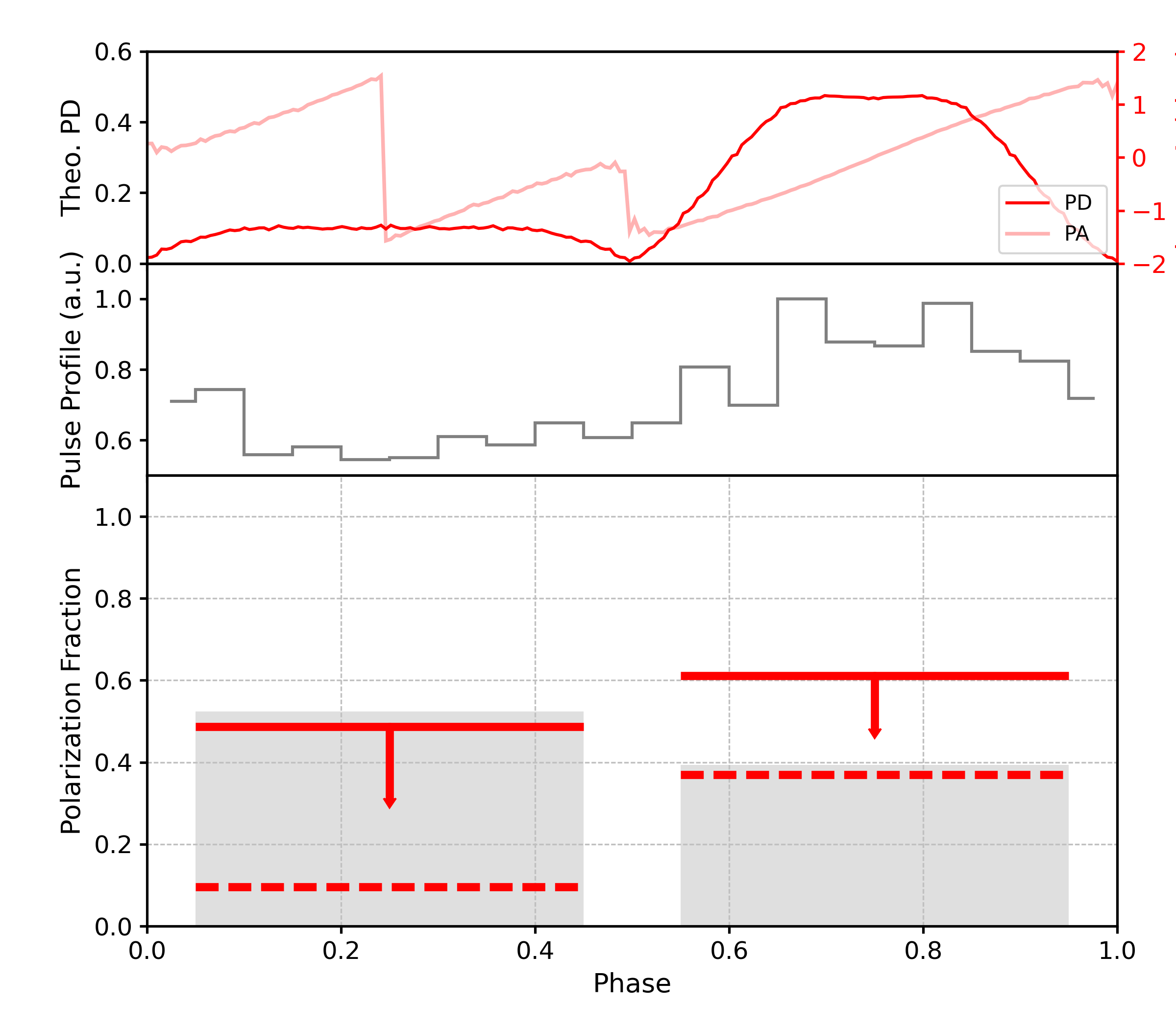}\\
    \includegraphics[width=0.45\linewidth, height=6cm]{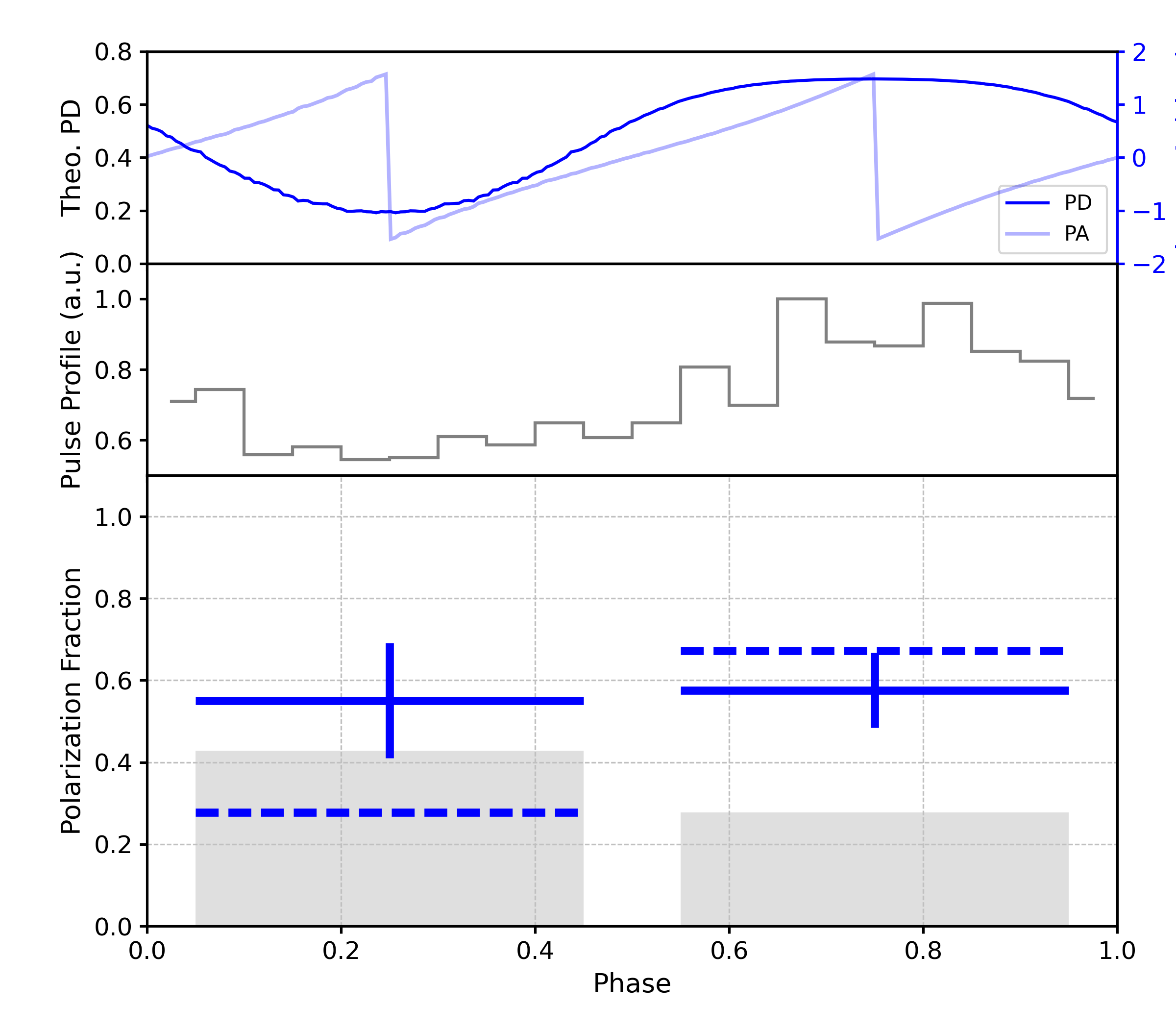}~~~~
    \includegraphics[width=0.45\linewidth, height=6cm]{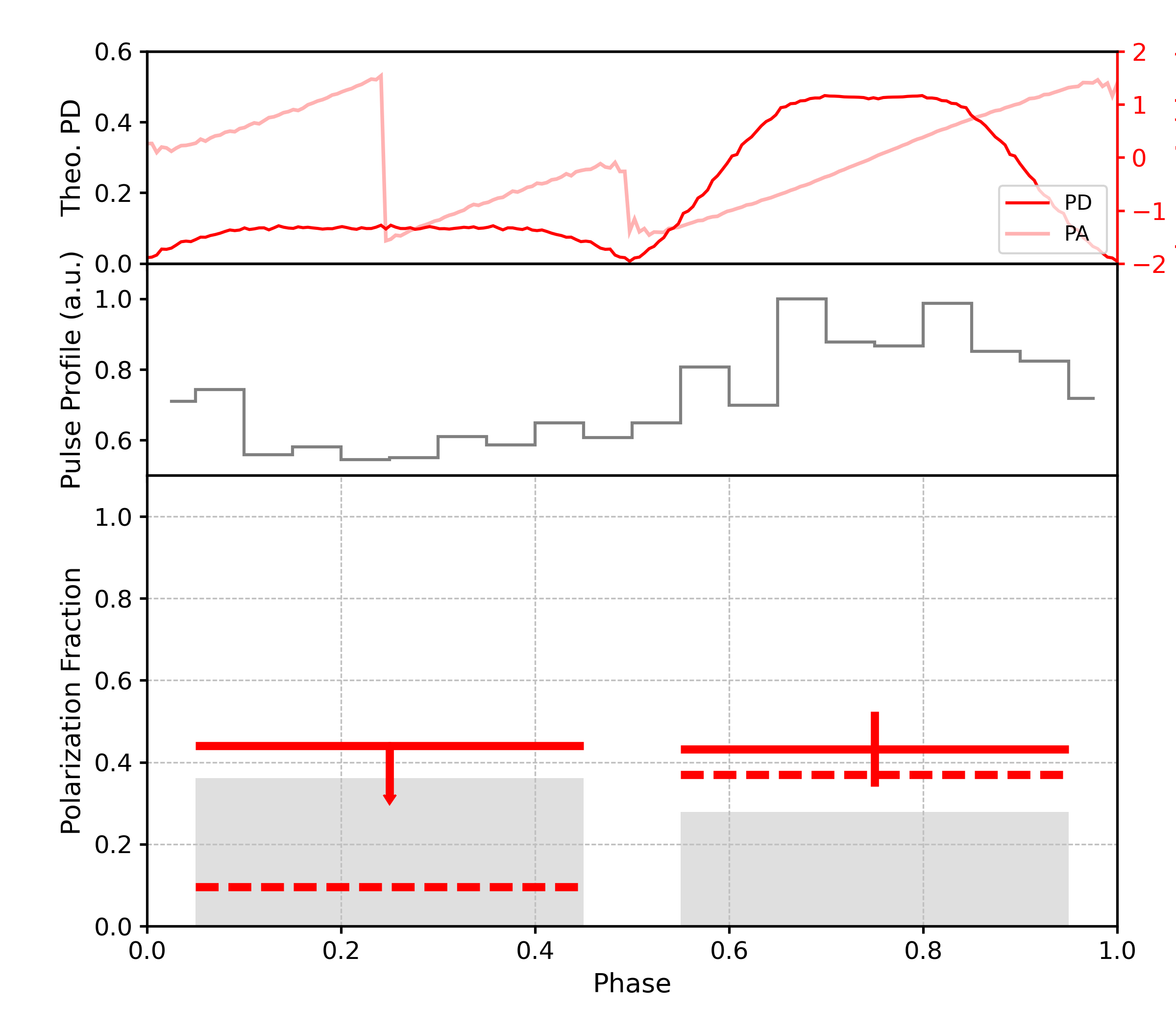}\\
    \caption{Top: Pulse profile of the simulated polarized and unpolarized component for the PER (left) and PAR (right) models. The real IXPE data pulse profile for \src\ is normalized to the maximum of the sum of the two components and overlaid for comparison. We make sure that the simulated observation carry the same statistics in the macro-phase bin [0.55-0.95] as the real one. Middle and bottom: phase-resolved polarization analysis of simulated IXPE exposures of $\times1$ (middle row) and $\times2$ (bottom row) of the current IXPE observation exposure ($\sim$ 1.254 Ms). In blue on the left column we illustrate the results for the PER model, while in red on right column we show the equivalent results for the PAR model. }
    \label{fig:unpolcomp}
\end{figure*}

The \texttt{ixpeobssim} simulation tools allow us to simulate the IXPE FoV pointing to a region of interest (ROI) centered on the target. In our case we used the coordinates of PSR J1723 RA=17h23m23.19s and Dec=-28d37m57.49s. In addition to the polarized component, we add an unpolarized phase-independent component, with the same spectrum as the polarized dominant component. The contribution of this component is such to match but not overshoot the maximum observed flux at the peak of the pulse profile (as illustrated in Fig.~\ref{fig:unpolcomp})\footnote{For scenarios in which the unpolarized emission is significantly higher than the baseline assumption ($\gtrsim\times$3), the expected polarization signal is correspondingly diluted, requiring substantially longer IXPE exposures to achieve a statistically significant detection.}. We included an instrumental background component in the ROI (through the class \texttt{xPowerLawInstrumentalBkg()}). These simulations with exposures are in order to generate enough statistics to confidently test our analysis pipeline.

Focusing on the results for the second phase bin (with the highest statistics) we highlight two key findings. First, given the exposure of the IXPE observation presented in this work, a polarized signal consistent with the PER model (as shown in the top-left panel of Fig.~\ref{fig:sims_exp}) would be detectable. Second, if the exposure were doubled, i.e., by repeating the observation with the same duration, we would reach the sensitivity required to detect a polarization signal as predicted by the PAR model. It is important to note that our simulations also account for the presence of an unpolarized emission component, providing a more realistic representation of the expected signal.

Before applying the phase-alignment pipeline, IXPE data are cleaned, barycenter-corrected, the source region and background region are selected according to standard procedures utilizing available tools within \texttt{ixpeobssim}, details of which can be found in appendix~\ref{subsec:ixpe}. Once the photons are aligned, we fold and bin the data in phase. For each phase macro-bin we evaluate the polarization through the \texttt{PCUBE} algorithm of \texttt{xpbin} available in the \texttt{ixpeobssim} package. 

\begin{figure*}
    \centering
    \includegraphics[width=0.5\linewidth]{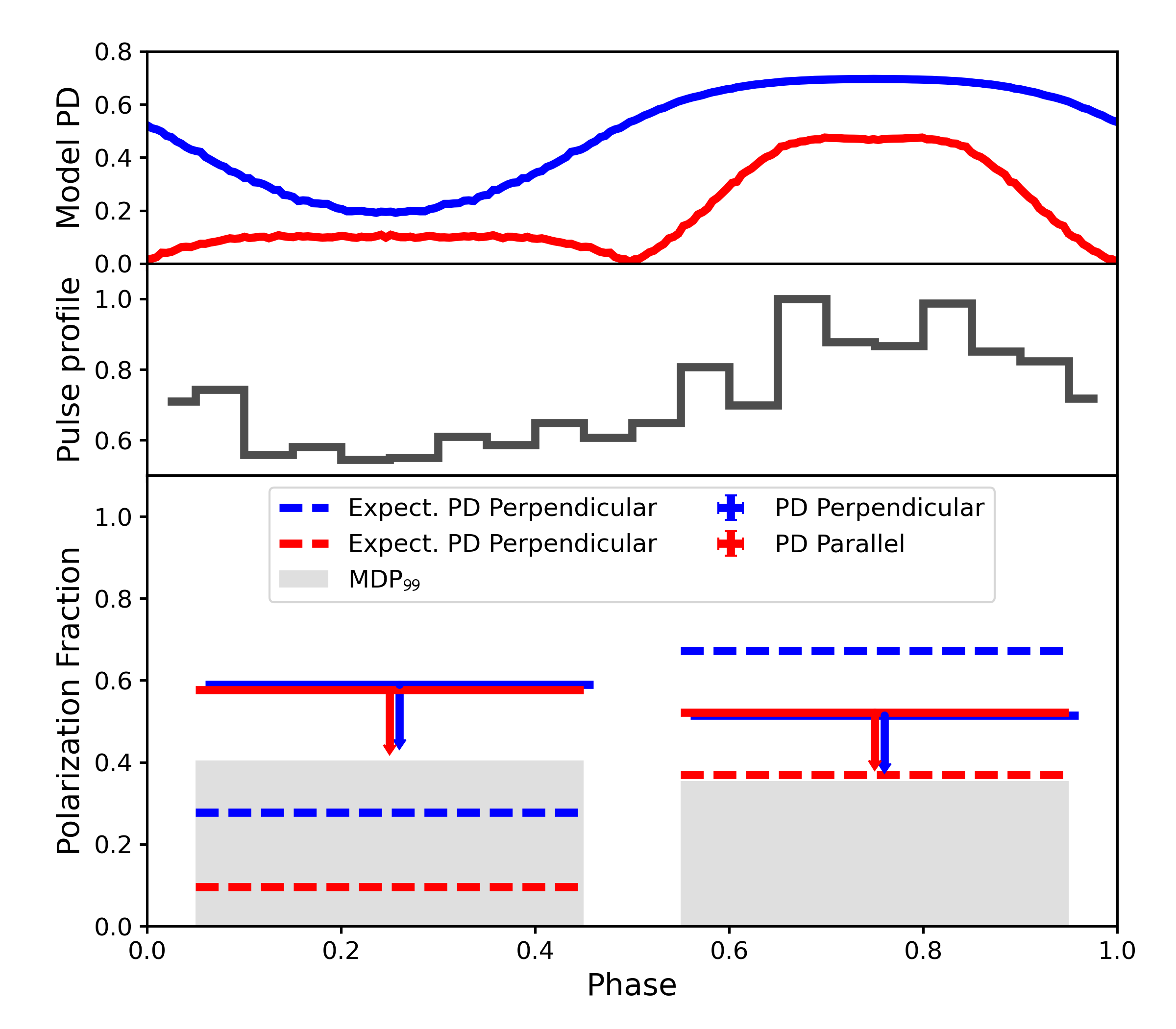}
    \caption{Results of the PCUBE analysis in the macro phase bins (0.05--0.45, 0.55--0.95) for the energy band 2--6 keV. The top panel shows the model-predicted polarization degree (PD) as a function of rotational phase for the perpendicular (blue) and parallel (red) configurations. The middle panel presents the pulse profile (arbitrary units), illustrating the different statistics in the two phase intervals used for polarization measurements. In the bottom panel, measured upper limits for the polarization degrees for both PAR (red) and PER (blue) configurations are shown. These are compared against the expected PD values from the models (dashed lines), while the shaded gray regions indicate the MDP$_{99}$.}
    \label{fig:PolResults}
\end{figure*}

The results of the phase-aligned PCUBE analysis on the real data are illustrated in Fig.~\ref{fig:PolResults}, in the macro-phase bins (0.05--0.45, 0.55--0.95) between 2--6 keV. Focusing on the 0.55--0.95 bin in which we have the peak of the statistics (${\rm MDP}_{99}=35\%$), we report a 99\% C.L. upper limit of about 51\%. This upper limit allows us to exclude a higher polarization, such as the one predicted by the PER model. In the first phase bin (0.05--0.45) the upper limit (of about 60\%) is not constraining due to the much lower PD predicted by both models.

\subsection{Spectropolarimetry}
 
The spectropolarimetric analysis was performed using both the multi-mission count spectra and the IXPE Stokes $Q$ and $U$ spectra. The latter were extracted from the Stokes-aligned dataset in the high-phase interval (0.55--0.95) for both the source and background regions, employing the \texttt{pha1q} and \texttt{pha1u} algorithms within the \texttt{ixpeobssim} \texttt{xpbin} routine. To account for the phase selection, the LIVETIME values in the PHA files were rescaled by a factor of 0.4, ensuring that XSPEC correctly interprets the count rates.  

We note that the IXPE observation is dominated by background above 7 keV. We therefore limit the polarization measurement in the 2--6 keV range.

The spectral model adopted here is identical to that used in the spectral-only fits described in Section~\ref{sec:spectroscopy}. The fit was carried out with the same set of free parameters as in the spectral analysis, augmented by two additional free parameters corresponding to a constant polarization degree and polarization angle . From this procedure, we obtain a $3\sigma$ upper limit on the PD of $53.7\%$, consistent with the results from the PCUBE analysis. We also tested a fit where only PD and PA were left free while fixing all other spectral parameters, which yielded identical results.

\section{Discussion, Summary and Conclusions} \label{sec:dicussion}

In this work, we analyzed the first IXPE observation of a redback spider pulsar, \src; we developed a model for the phase-modulated polarized X-ray emission from this system, driven by the rotation of the pulsar, and we identified two distinct physical scenarios—parallel (PAR) and perpendicular (PER)---characterized by the orientation of the magnetic field lines with respect to the bulk flow of particles. We devised a novel method to align the photons’ electric vector position angle to a predicted, phase-dependent direction, allowing us to probe a varying polarization angle across phase bins while assuming a phase-dependent polarization degree, and we validated this alignment technique through dedicated simulations, assessing the expected polarization signature under increased photon statistics.

We placed upper limits on the polarization degree in two phase-resolved bins; however, in the highest-statistics phase bin, the predicted PDs for the PAR and PER models are sufficiently distinct to exclude the PER scenario at the current sensitivity level. As demonstrated by the simulations above, a high PD over $60\%$ the phase bin 0.55--0.95 in the ``perpendicular'' case would have been detected in our exposure. Such a case is deliberately extreme and corresponds to particles which radiate in plasmoid-like flux rope structures along the shock. Where particles radiate compared to where they are accelerated is an open question in spider binaries, including the thickness of the zone of activity. Flux freezing in a MHD-like bulk flow naturally favors a global structure which is more akin to the ``parallel” case. Departures from MHD, and the presence of ions in the shock, however could significantly change the shock structure geometry. The amplitude of field turbulence which may depolarize signals (and at which spatial scales it is present), is also an open question.

The ``parallel” and “perpendicular” cases and their constraints are model-dependent insofar as the electic vector polarization angle (EVPA) evolution is stipulated and manifestly quasi-linear by construction. However, as noted in appendix~\ref{app:align2}, we also considered model-independent EVPA rotations for linear and fully-general Fourier series models for EVPA evolution. These again yielded upper limits, though with some hint (whose statistical significance is difficult to rigorously quantify) of preference clockwise rotation on the sky in slope of EVPA linear models on the data (bottom panel of Figure~\ref{fig:sim_data_model_results}). We caution against overinterpretation of such a preference, as these features were also seen in some cases of unpolarized data simulations (middle panel of Figure~\ref{fig:sim_data_model_results}). These advanced techniques highlight the possibility to detecting subtle features possibly already present in the data not revealed by standard IXPE analysis methods. Note that this technique ought to be applied to IXPE data of other sources whose intrinsic PA variation is \textit{a priori} not known but is strictly periodic at a known period.

Through simulations, we also show that doubling the exposure time would enable a detection of a polarization signal consistent with the PAR model. Given the demonstrated long-term stability of \src\ across multiple X-ray observations spanning over a decade, increased IXPE exposure and combining the datasets would be a feasible and effective strategy to reach the required sensitivity.

In conclusion, \src\ could benefit from additional IXPE observations or observations from more sensitive upcoming X-ray polarimeters (e.g. the Enhanced X-ray Timing and Polarimetry \citep[eXTP][]{2025SCPMA..6819502Z}). Such observations would significantly enhance the ability to detect and characterize the phase-dependent polarization signal and ultimately constrain the particle acceleration and plasma structures proximate to the intrabinary shock.

\begin{acknowledgments}
We thank George Younes, Kostas Kalapotharakos, and Jorges Cortes for helpful discussions. Z.~W. and H.Z. acknowledge support by NASA under award numbers 80GSFC21M0002 and 80GSFC21M0006. 
Z.~W., M.~N, and S.~B. were supported in part by NASA IXPE General Observer program grant 80NSSC25K7234. H.Z. also acknowledges support by NASA under award number 80NSSC24K1173. J.H. acknowledges support from NASA under award number 80GSFC21M0002. Simulations were carried out on the NASA Pleiades cluster and NERSC Perlmutter cluster. 
This paper employs a list of Chandra datasets, obtained by the Chandra X-ray Observatory, contained in the Chandra Data Collection (CDC) 'collection\_id'~\href{https://doi.org/10.25574/cdc.539}{[DOI:10.25574/cdc.539]}. This research has made use of data and software provided by the High Energy Astrophysics Science Archive Research Center (HEASARC), which is a service of the Astrophysics Science Division at NASA/GSFC and the High Energy Astrophysics Division of the Smithsonian Astrophysical Observatory, and the NuSTAR Data Analysis Software (NuSTARDAS), jointly developed by the ASI Space Science Data Center (SSDC, Italy) and the California Institute of Technology (Caltech, USA). The work presented is based in part on observations with \textit{XMM-Newton}, an ESA Science Mission with instruments and contributions directly funded by ESA Member states and NASA. This research has made use of data from the NuSTAR mission, a project led by Caltech, managed by the Jet Propulsion Laboratory, and funded by NASA. This work made use of Neil Gehrels Swift Observatory data supplied by the UK Swift Science Data Centre at the University of Leicester. This research has made use of data obtained from the \chandra~ Data Archive and software provided by the \chandra~ X-ray Center (CXC) in the application package CIAO. This work has relied on the NASA Astrophysics Data System. 

\end{acknowledgments}

\vspace{5mm}
\facilities{CXO, IXPE, NuSTAR, Swift-XRT, XMM}

\software{ixpebssim \citep{2022SoftX..1901194B}, HEASoft \citep{2014ascl.soft08004N}, SAS \citep{2004ASPC..314..759G}, NuSTARDAS}

\bibliography{bibfile}{}
\bibliographystyle{aasjournalv7}

\appendix

\section{Data reduction}
\label{app:data}

\subsection{IXPE}
\label{subsec:ixpe}

The Imaging X-ray Polarimetry Explorer \citep[IXPE;][]{IXPEPreLaunch} observed \src~ in two sections, starting on 2024-08-20T18 and 2024-09-01T06, respectively, for a total effective exposure (livetime) of 1254335.9 s. The two observations have been merged into one OBSID (03006799). The data have been cleaned from the majority of the misidentified charge cosmic rays utilizing the background rejection procedure outlined in \citet{DiMarcoCut}. We look for possible spurious flares caused by the solar activity and identified only two 60 seconds long intervals where the rate exceeds 99\% quintile threshold. In the whole FoV, only a few high-energy photons ($>4$ keV) fall in these intervals and they do not affect the results once the region selection around the binary is applied. 

\begin{figure}
    \centering
    \includegraphics[height=7cm]{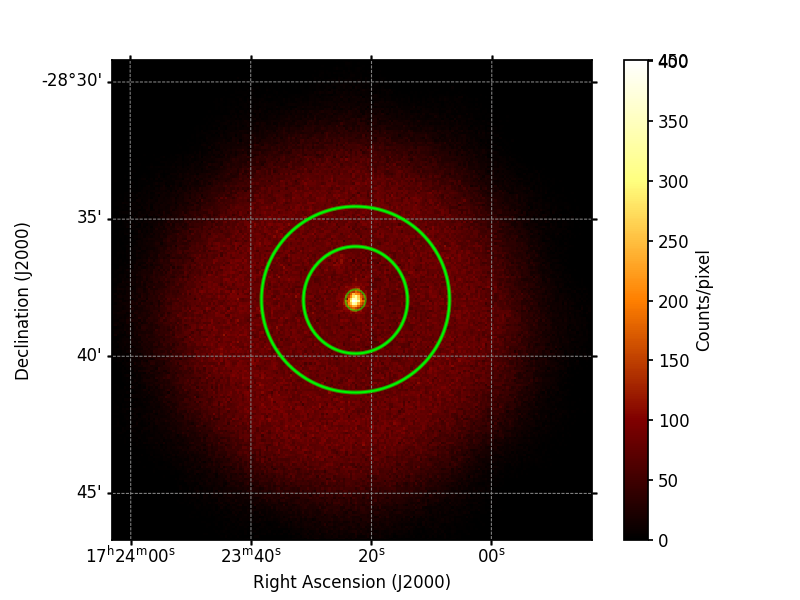}
    \includegraphics[height=7cm]{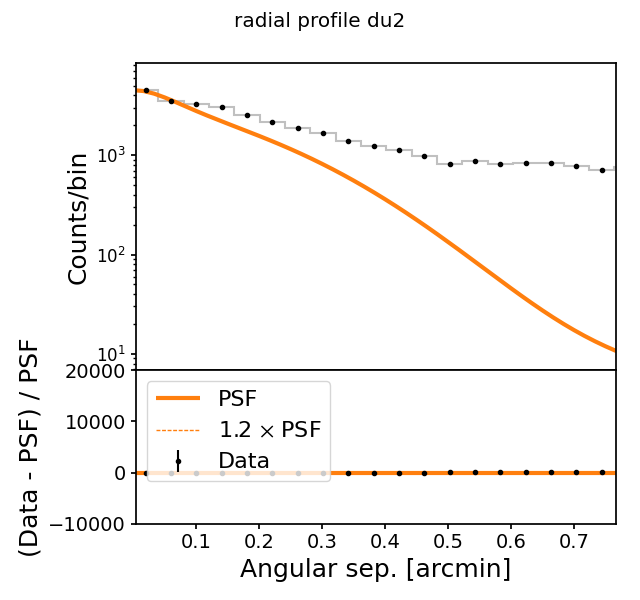}
    \caption{\textit{Left:} Counts map of the IXPE observation (all DUs). in green we mark the selection regions for the source (think innermost circle) and the background (annular region between the two thinker circles). \textit{Right:} Radial profile centered on the peak-emission in the field of view of DU2 (as an example, DU1 and DU3 look similar and hold the same information). The peak emission has been obtained with a 2D-Gaussian fit for each DU.}
    \label{fig:radial_profile}
\end{figure}

We use \texttt{ixpeobssim} (v31.0.2) to select (\texttt{xpselect}) and analyze (\texttt{xpbin}) the data. We define the source region \textit{a priori}, by selecting all the photons within the peak emission from the pulsar with radius equal to the average FWHM of the IXPE telescope units (about 30$''$). The study of the radial profile obtained from the observations confirmed that within the selected region we can expect to have mostly photons from the central point source, while at larger radii the contribution from the background starts to be dominant, contributing significantly to the background of our measurement (see Fig.~\ref{fig:radial_profile}). The astrophysical background, used for background subtraction, is extracted from an annular region around the source with inner and outer radius of 2$'$ and 3.5$'$. 

IXPE I, Q and U spectra have been obtained using the \texttt{xpbin} routine of \texttt{ixpeobssim}, selecting the \texttt{pha1, pha1q, pha1u} algorithms respectively. The source and background spectra reveal that the background becomes dominant beyond $\sim$ 6 keV, we therefore limit the IXPE polarization data analysis to the 2--6 keV energy band.

\subsection{NuSTAR}
\src\ was targeted with the Nuclear Spectroscopic Telescope Array (NuSTAR; \citealt{2013ApJ...770..103H}) starting on 2015 October 15 (ObsID 30101043002) in a 81.9 ks deadtime-corrected on-source exposure. The event data were collected over a total elapsed time of 165.7 ks, covering slightly over three full orbital cycles of the binary. The event data were first processed using the \texttt{nupipeline} script in NuSTARDAS, while  spectra and barycentered, background-subtracted light curves were produced with the \texttt{nuproducts} task. For the light curve analysis, the time series from both focal plane modules (FPMA and FPMB) were stacked.

\subsection{XMM-Newton}
PSR J1723$-$2837 was targeted by XMM-Newton on 2011 March 3 in a single continuous 56.4 ks exposure. For the variability and spectroscopic analyses we only consider the European Photon Imaging Camera (EPIC) pn data set since a substantial portion of the MOS1/2 exposures are strongly affected by background flaring.
The XMM-Newton observations were processed and reduced using the with the Science Analysis Software (SAS) version \texttt{xmmsas\_20211130\_0941-20.0.0} by running the unprocessed (ODF) data products through the \texttt{epproc} pipeline. The event redistribution matrix and the auxiliary response file for the observation were generated using the \texttt{rmfgen} and \texttt{arfgen} tasks in SAS. The correct scalings of the areas of the source regions used to extract the source and background spectral files were computed with the \texttt{backscale} tool.

\subsection{\chandra~ X-ray Observatory}
The \chandra~ ACIS data products were extracted with the aid of CIAO version 4.15 \citep{2006SPIE.6270E..1VF} and the accompanying calibration database CALDB 4.10.4.
The source and background spectra, along with the ancillary response products, were generated using the \texttt{specextract} task in CIAO. The barycentering of the \chandra~ source events was done using the \texttt{axbary} task in CIAO assuming the DE405 JPL solar system ephemeris. 

\subsection{Swift-XRT}

Swift-XRT observed \src\ for 8.9 ks between 2024-08-23 06:30:02 and 2024-08-28 23:47:59 UT (operating in Photon Counting mode), to provide simultaneous coverage to the IXPE observations and ensure the system had not changed when compared to archival observations/analysis. 
%The XRT was operated in Photon Counting (PC) mode.
To process the archival data and produce the phase-averaged spectrum, we utilized the {\tt build Swift-XRT products} tool\footnote{\url{https://www.swift.ac.uk/user_objects/}}.
This applies all standard calibrations/filtering, and extracts spectra using the standard tools ({\tt XSELECT}), as described by \cite{2009MNRAS.397.1177E}.

\startlongtable
\begin{deluxetable}{ccccc}
\centering
\tabletypesize{\scriptsize}
\tablecolumns{5} 
\tablewidth{0pt} 
\tablecaption{Log of X-ray Observations of \src\ \label{tab:xraylog}}
\tablehead{
\colhead{Telescope}  & \colhead{Instrument} & \colhead{ObsID} & \colhead{Start Time} & \colhead{Exposure} \\
 \colhead{} & \colhead{} & \colhead{} & \colhead{(UTC)} & \colhead{(ks)}   }
\startdata
IXPE       & DU1        & 03006799       & 2024-08-20 23:59:07 &  1254.3 \\	
           & DU2        &                &                         &  1254.3\\
           & DU2        &                &                         &  1254.1 \\
\hline
XMM-Newton & EPIC pn    & 0653830101	 &  2011-03-03 13:00:53  &	   56.4          \\
\hline
\chandra~    & ACIS-S     & 13713          & 2012-07-11 05:21:08   &     55.1            \\
\hline
NuSTAR     & FPMA/FPMB  & 30101043002    & 2015-10-15 05:36:06   &     	81.9            \\
\hline
	Swift      & XRT        	&	00031643001	&	2010-03-05 17:11:04	&	0.8	\\
	&		&	00031643002	&	2010-03-10 00:05:00	&	5.8	\\
	&		&	00092033001	&	2015-01-29 03:38:59	&	1.4	\\
	&		&	00092033002	&	2015-02-26 09:00:58	&	1.0	\\
	&		&	00092033003	&	2015-03-26 10:45:59	&	0.9	\\
	&		&	00092033006	&	2015-05-02 12:14:59	&	0.8	\\
	&		&	00092033007	&	2015-05-21 09:46:59	&	0.8	\\
	&		&	00031643003	&	2015-05-25 14:35:58	&	0.3	\\
	&		&	00092033008	&	2015-06-18 06:44:59	&	1.0	\\
	&		&	00092033009	&	2015-07-16 13:03:58	&	0.8	\\
	&		&	00031643004	&	2015-07-20 11:13:58	&	0.8	\\
	&		&	00092033010	&	2015-08-13 08:27:58	&	1.2	\\
	&		&	00092033011	&	2015-09-10 14:45:57	&	0.8	\\
	&		&	00031643005	&	2015-09-14 01:55:58	&	1.0	\\
	&		&	00092033012	&	2015-10-08 11:22:57	&	0.9	\\
	&		&	00031643006	&	2015-10-15 04:29:53	&	2.0	\\
	&		&	00081656001	&	2015-10-15 11:28:58	&	1.9	\\
	&		&	00031643007	&	2015-10-16 02:52:57	&	1.5	\\
	&		&	00031643008	&	2015-10-21 04:08:58	&	1.8	\\
	&		&	00031643009	&	2015-10-27 00:33:57	&	0.9	\\
	&		&	00031643010	&	2015-10-29 00:37:58	&	0.6	\\
	&		&	00031643011	&	2016-01-29 15:11:58	&	1.0	\\
	&		&	00031643012	&	2016-02-02 16:36:58	&	1.0	\\
	&		&	00031643013	&	2016-03-25 22:30:58	&	0.1	\\
	&		&	00031643014	&	2016-03-31 22:07:58	&	1.0	\\
	&		&	00092227001	&	2016-05-07 11:15:58	&	1.0	\\
	&		&	00092227002	&	2016-05-31 11:08:02	&	0.2	\\
	&		&	00092227003	&	2016-06-03 14:15:58	&	0.8	\\
	&		&	00092227004	&	2016-06-24 06:04:58	&	1.0	\\
	&		&	00092227005	&	2016-07-19 01:25:57	&	1.1	\\
	&		&	00092227006	&	2016-08-10 02:29:58	&	0.9	\\
	&		&	00092227007	&	2016-09-04 15:04:57	&	1.0	\\
	&		&	00092227008	&	2016-09-28 03:44:58	&	1.1	\\
	&		&	00092227009	&	2016-10-22 14:40:58	&	1.0	\\
	&		&	00092411001	&	2017-01-30 06:27:56	&	1.0	\\
	&		&	00092411002	&	2017-02-27 10:17:57	&	1.1	\\
	&		&	00092411003	&	2017-03-27 19:19:57	&	1.0	\\
	&		&	00092411004	&	2017-04-24 02:30:57	&	1.0	\\
	&		&	00092411005	&	2017-05-22 19:16:57	&	0.9	\\
	&		&	00092411006	&	2017-06-19 05:46:57	&	1.0	\\
	&		&	00092411007	&	2017-07-17 10:00:57	&	1.0	\\
	&		&	00092411008	&	2017-08-14 05:49:57	&	1.0	\\
	&		&	00092411009	&	2017-09-10 16:24:56	&	1.0	\\
	&		&	00092411010	&	2017-10-09 04:10:57	&	1.0	\\
	&		&	00094150001	&	2019-02-15 10:39:36	&	1.3	\\
	&		&	00094150002	&	2019-03-15 20:44:36	&	0.9	\\
	&		&	00094150003	&	2019-04-15 22:24:35	&	1.1	\\
	&		&	00094150004	&	2019-05-15 00:51:36	&	0.9	\\
	&		&	00094150006	&	2019-07-16 13:55:34	&	1.0	\\
	&		&	00031643015	&	2019-07-28 22:08:36	&	0.2	\\
	&		&	00094150007	&	2019-08-15 14:09:35	&	0.9	\\
	&		&	00094150008	&	2019-09-15 22:24:35	&	0.8	\\
	&		&	00094150009	&	2019-10-15 00:12:35	&	1.0	\\
	&		&	00095613001	&	2020-04-25 20:01:36	&	1.0	\\
	&		&	00095613002	&	2020-07-25 22:05:35	&	1.0	\\
	&		&	00095613003	&	2020-10-25 00:01:35	&	0.8	\\
	&		&	00095613004	&	2021-01-29 01:20:34	&	0.8	\\
	&		&	00095801001	&	2021-05-22 20:15:35	&	0.9	\\
	&		&	00095801002	&	2021-08-14 06:33:48	&	0.6	\\
	&		&	00095801003	&	2021-10-26 00:34:35	&	0.9	\\
	&		&	00096542001	&	2022-05-07 11:53:35	&	1.0	\\
	&		&	00096542002	&	2022-07-30 04:20:35	&	0.9	\\
	&		&	00096542003	&	2022-10-22 18:24:36	&	0.8	\\
	&		&	00096542004	&	2023-01-30 01:54:35	&	0.8	\\
    &      & 00031643016\tablenotemark{a}	 & 2024-08-23 06:28:55  &     5.2  \\
           &            &  00031643017\tablenotemark{a}   & 2024-08-25 04:10:57	 &  1.2   \\
           &            &  00031643018\tablenotemark{a}   & 2024-08-27 23:58:56	 &  2.6   \\
\enddata
\tablenotetext{a}{Swift exposures contemporaneous with the IXPE observation.}
\end{deluxetable}

\section{Polarization Angle phase Alignment}
\label{app:align}
As \src\ is a compact binary with $P_{\rm b} = 14.8$ hr, it rotates on the sky several orbits ($\sim 24$ times) during the IXPE exposure. In a binned phase-folded analysis, rotation of binary (i.e. fast PA evolution) on the sky may depolarize the accumulated signal and hamper detection of intrinsically polarized emission. To mitigate such depolarization, we have developed a custom alignment EVPA procedure and tested its efficacy with simulations.

In section \ref{app:align1} we provide the mathematical derivation of the phase alignment procedure. Additionally, in section \ref{app:align2} we present our attempts to gain insights on the phase dependence of the polarization angle via a maximum likelihood fit.

\subsection{Phase-alignment Formalism}
\label{app:align1}

\begin{figure}
    \centering
    \includegraphics[width=\linewidth]{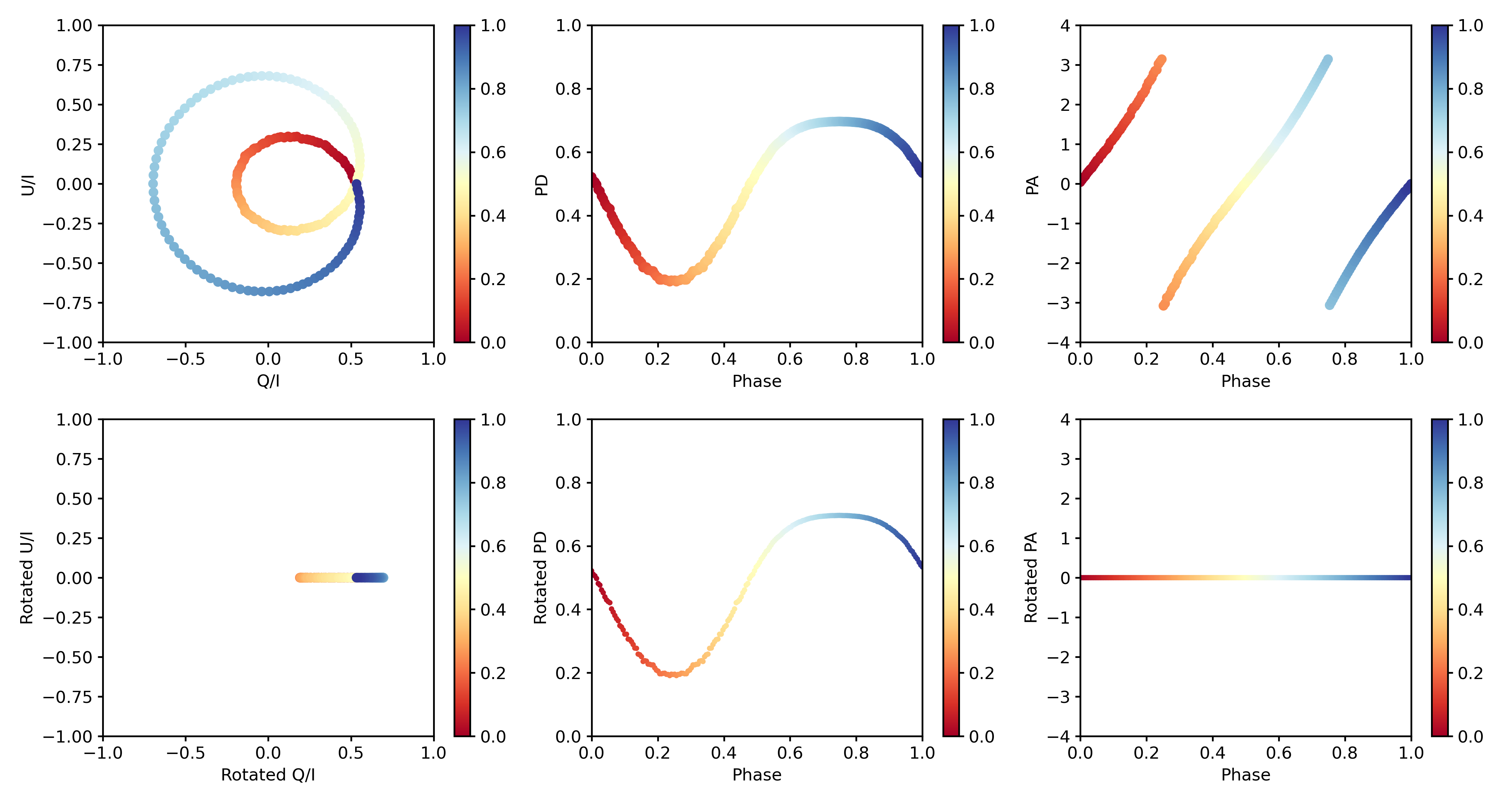}
    \caption{Effect of the EVPA alignment on the phase-dependent polarization properties of the PER model. The top row shows the unaligned Stokes chart (left), the polarization degree (middle), and the polarization angle(right). The color bar illustrates the phase evolution from red to blue. The bottom row illustrates the same quantities after applying the alignment procedure using the input PA of the model. The alignment effectively rotates the Stokes vectors maintaining the polarization degree at each phase but rotating all PA to be aligned with zero degrees. This procedure allows for binning wider in phase without being depolarized by the fast angle rotation of the binary on the sky.}
    \label{fig:sim_rot}
\end{figure}

Figure \ref{fig:sim_rot} illustrate the effect of the Stokes parameter rotation to align all the angles in each phase to the same (zero) angle. Mathematically, this operation consists in the following Stokes-vector rotation:
\begin{equation}
    \begin{bmatrix}
    q_{\rm aligned}\\
    u_{\rm aligned}
    \end{bmatrix} = 
    \begin{pmatrix}
    \cos(2\phi) & -\sin(2\phi)\\
    \sin(2\phi) & \cos(2\phi)
    \end{pmatrix}
    \begin{bmatrix}
    q_0\\
    u_0
    \end{bmatrix}
\end{equation}
where $\phi$ is the PA provided by the model, and $q_0$ and $u_0$ are the initial values.

\begin{figure}[t]
    \centering
    \includegraphics[width=0.45\linewidth]{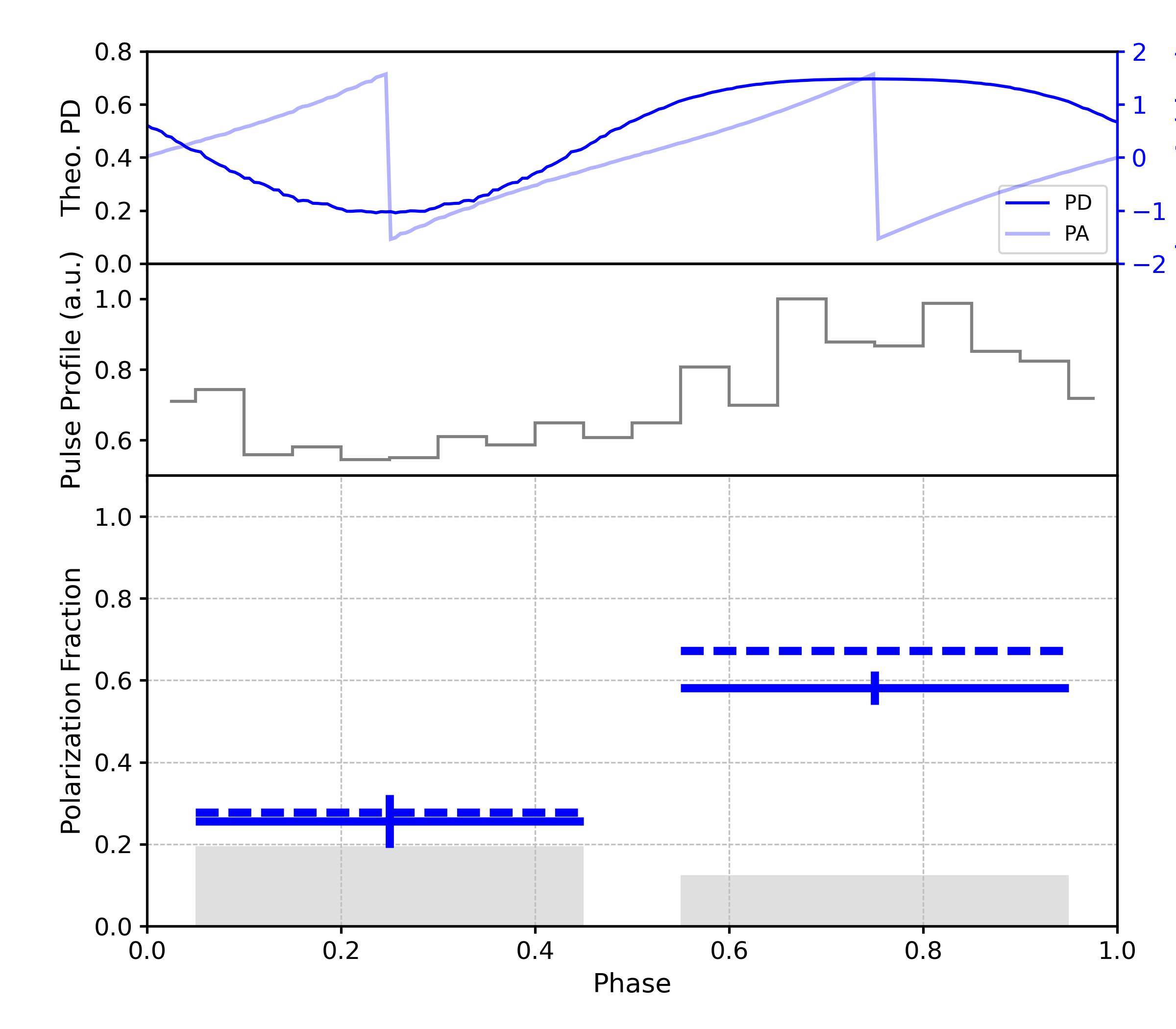}~~~~
    \includegraphics[width=0.45\linewidth]{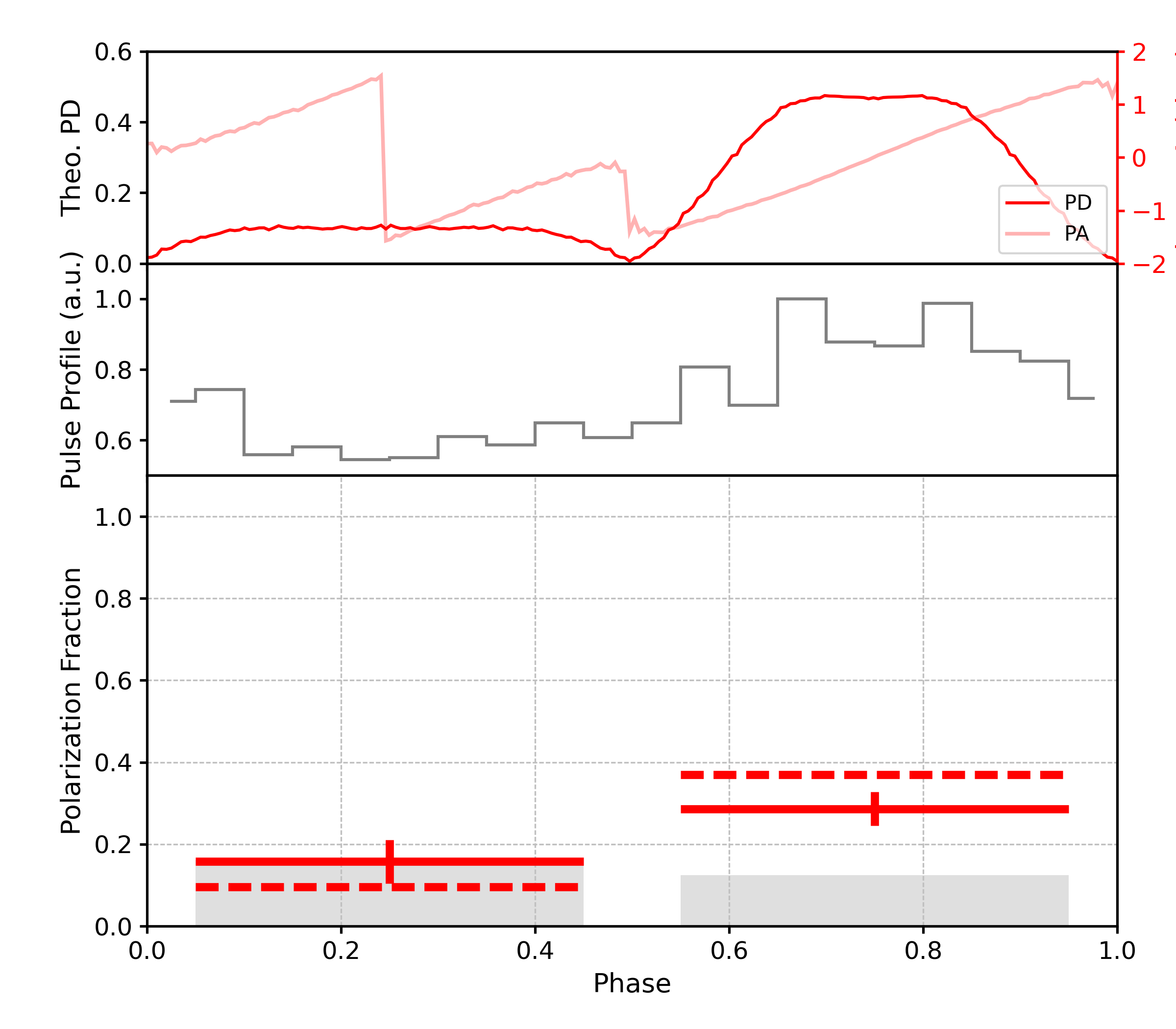}
    \caption{Phase-resolved polarization analysis of simulated IXPE exposures of $\times10$ (bottom) the real IXPE observation ($\sim$ 1.254 Ms). In blue on the left column we illustrate the results for the PER model, while in red on right column we show the equivalent results for the PAR model. The gray shaded areas mark the MDP$_{99}$, while the dashed lines mark the true (simulated) average PD values.}
    \label{fig:sims_exp}
\end{figure}

However, in the context of the photon-by-photon rotation that we want to apply to IXPE data, we need to account for the fact that the photon-by-photon Stokes parameters are defined as 
\begin{equation}
    q^{\gamma} = 2 \cos(2\varphi), ~~ \\
    u^{\gamma} = 2 \sin(2\varphi)
\end{equation}
for each photon with Stokes parameters $q^{\gamma}$ and $u^{\gamma}$ and associated phase we derive the model's predicted Stokes $q_m$ and $u_m$ at the same phase and the direction we want to align with as:

\begin{equation}
    \vec{D}_m\frac{1}{\sqrt{q_m^2 + u_m^2}} 
    \begin{bmatrix}
    q_m\\
    u_m
    \end{bmatrix}
\end{equation}
The alignement of the photons Stokes parameters is then computes as

\begin{equation}
    q_{\rm aligned} = \frac{(q^{\gamma} q_m + u^{\gamma} u_m)}{\sqrt{q_m^2 + u_m^2}}, ~~ \\
    u_{\rm aligned} = \frac{(u^{\gamma} q_m - q^{\gamma} u_m)}{\sqrt{q_m^2 + u_m^2}}
\end{equation}
The new aligned stokes parameters are then saved to a new file and analyzed in phase bins.
The sequence of Fig.~\ref{fig:sims_exp} illustrates the results of the phase binned analysis adopting the same binning as for the real observation (but for an ideal case of $\times10$ the exposure). 

\begin{figure}
    \centering
    \includegraphics[width=0.45\linewidth]{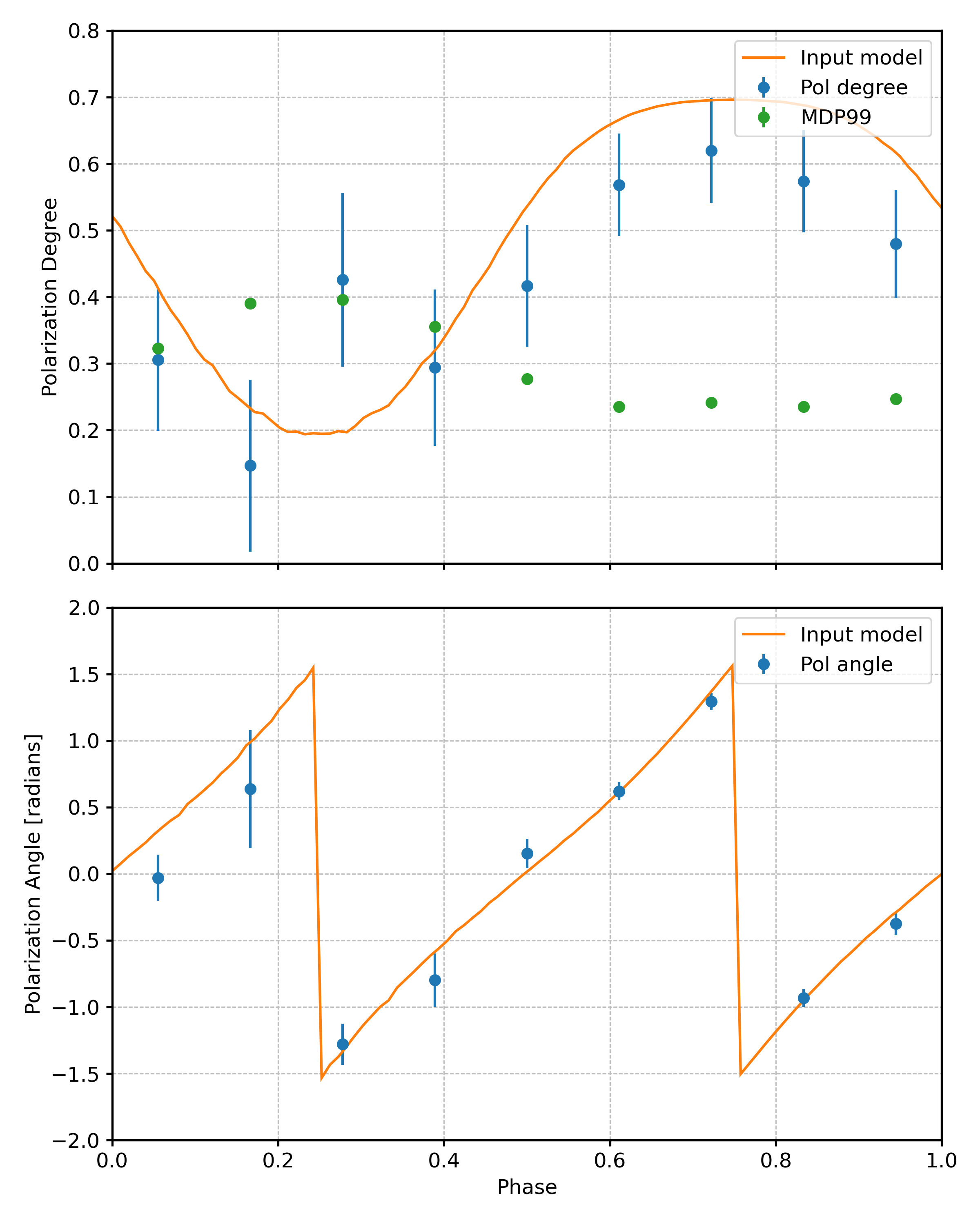}~~~~
    \includegraphics[width=0.45\linewidth]{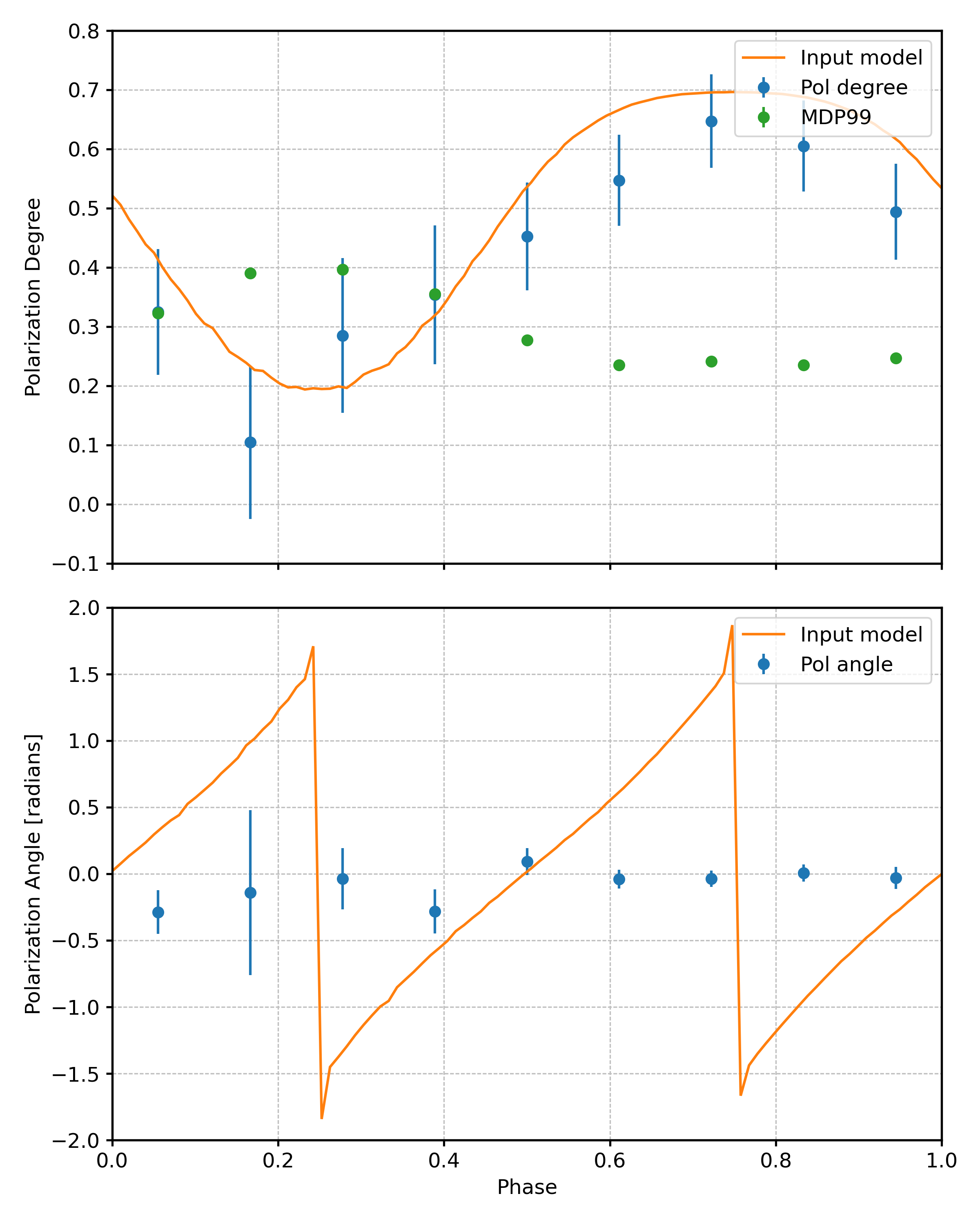}
    \caption{Recovered phase dependent PD and PA for simulated IXPE observation (assuming the PER model) for the $\times10$ exposure case. On the left we show the pre-alignment results, while on the right we illustrate the results after the alignment. Note that in finer phase bins the depolarization effect due to the PA rotation is less strong and allows for th erecovery of the polarized signal with only a minor bias. The alignment becomes important with larger phase bins and lower statistics. }
    \label{fig:simx10_10bins}
\end{figure}

In Fig.~\ref{fig:simx10_10bins} we illustrate the analysis more finely binned in phase for the $\times10$ exposure case. In particular we show the effect of the phase-alignment procedure by reporting the recovered PD and PA before and after the application of the alignment for the PER model (an equivalent plot can be derived for the PAR model).

\subsection{Searching for PA-phase model with maximum likelihood approach}
\label{app:align2}
We explored the possibility of inferring the phase-dependence of the polarization angle to gain some insights on the best physical model that could describe the observation. The goal is to probe possible phase-dependent variations in the polarization angle, with the broader aim of constraining the underlying physical model that best describes the observations. Rather than attempting to measure the PA directly, the strategy is to maximize the ratio of PD to its statistical uncertainty ($\Delta \mathrm{PD}$) by allowing the PA to vary in phase according to a given functional form. In practice, the fit is performed by maximizing the following likelihood ($\mathcal{L}$):
\begin{equation}
    \mathcal{L} = \frac{\rm PD}{\Delta \rm PD} = \frac{\rm PD}{\sqrt{\frac{2 - \bar{\mu}^2 \rm PD^2}{\bar{\mu}^2(N - 1)}}}
\end{equation}
where $N$ is the total number of events, $\bar{\mu}$ is the average modulation factor and $\Delta$PD is calculated 
according to \citet{2015APh....68...45K}. At each step of the fitting process, PD is evaluated from the event Stokes parameters $q$ and $u$ aligned with the current PA model as a function of phase:
\begin{equation}
    \mathrm{PD} = \frac{\sqrt{(\sum q_{\rm aligned})^2 + (\sum u_{\rm aligned})^2}}{N}.
\end{equation}
We tested two sets of models: 1) a piecewise linear model, mimicking a constant rotation, for which we vary the number of possible breaks and segments allowed (this is to account for the `jumps' in PA due to the discontinuity of the parameter space); 2) a Fourier model, for which we vary the order of expansion (i.e. the number of harmonics in a Fourier series).

\begin{figure}
    \centering
    \includegraphics[width=0.45\linewidth]{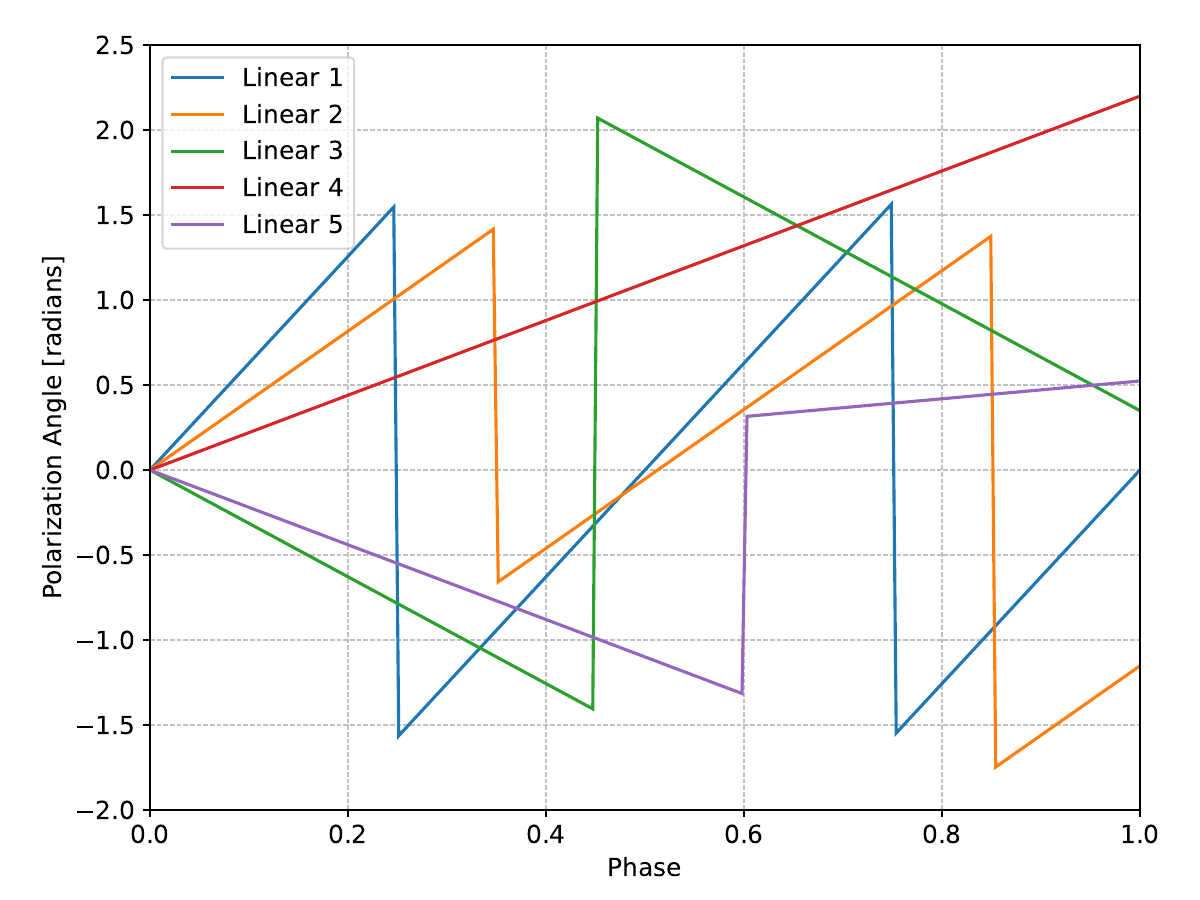}~~~~
    \includegraphics[width=0.45\linewidth]{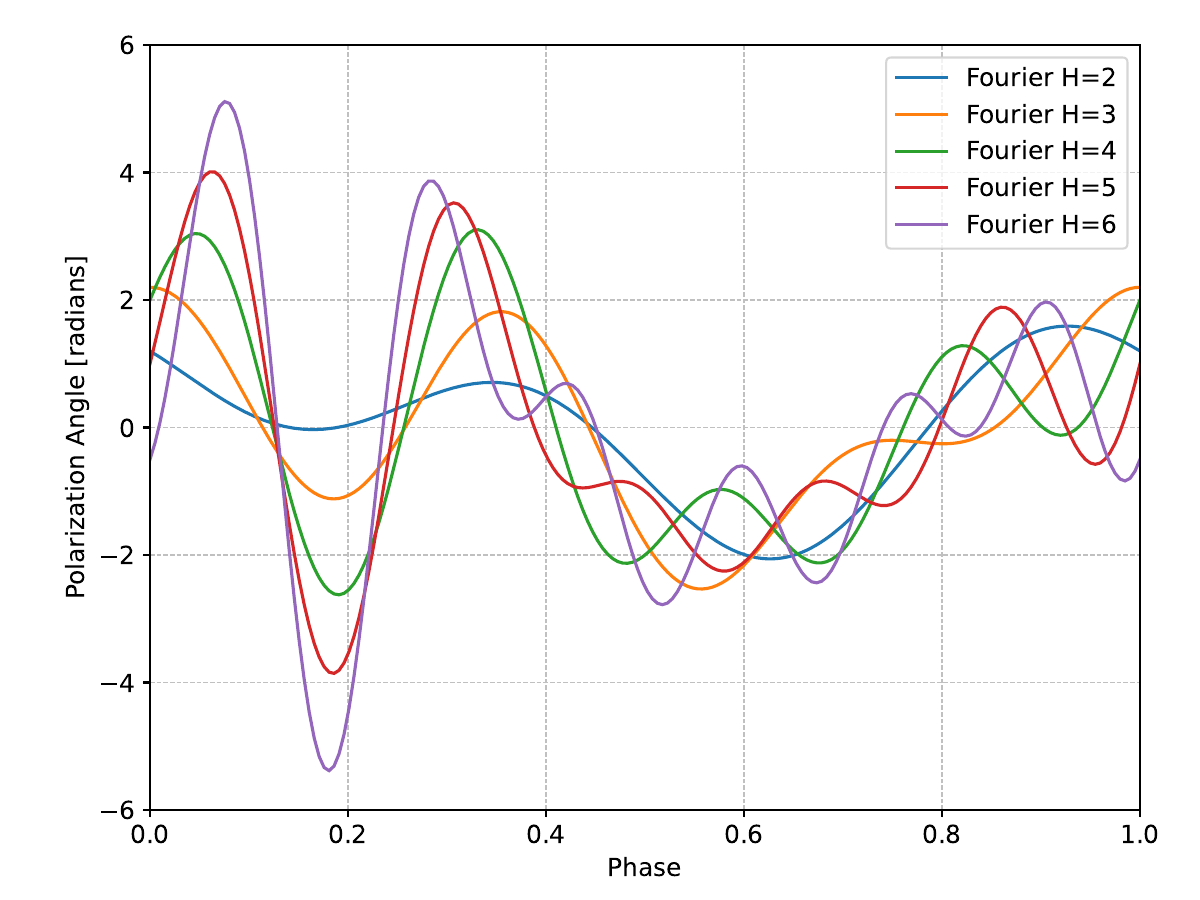}
    \caption{Example of linear (left) and Fourier functional forms (right) used to model the PA change in phase. The set of linear models represented are: one slope, two fixed jumps (blue), one slope, two free jumps (orange), one slope, one free jump (green), one slope, no jumps (red), two slopes, one free jump (purple). The set of Fourier models plotted are, in order of increasing number of harmonics: $H=2$ (blue), $H=3$ (orange), $H=4$ (green), $H=5$ (red), $H=6$ (purple).}
    \label{fig:lin_fou_models}
\end{figure}

The linear models consist of a series of connected linear segments as a function of the PA, which is hence assumed to evolve linearly in phase with constant EVPA rotation. The number of segments, their slope, and the presence or absence of discontinuous jumps between them are varied, resulting in a range of models with different numbers of free parameters. The left side of Figure~\ref{fig:lin_fou_models} shows some examples of linear models used in this analysis. For instance, models with a single slope and two unconstrained phase jumps were tested against configurations where either the slope or the jump amplitudes were fixed to physically motivated values (e.g., $\pm2\pi$ for the slope or $\pm$90° for the jumps). These tests were performed on both unpolarized simulations and real data, with some configurations allowing up to three linear segments and two phase discontinuities. This type of analysis is not sensitive to any global shift of PA. Thus, the intercept of the linear model at $\phi=0$ was fixed to 0, and treated as a nuisance parameter.

The second class of models is based on Fourier decomposition. Here, the PA is described as a Fourier series with the following functional form:
\begin{equation}
    \mathcal{F}(\phi) = a_0 + \sum_{n=1}^{H} a_n \cos(2\pi \, n\phi) + b_n \sin (2\pi \, n\phi)
\end{equation}

where the number of harmonics $H$ is varied. This effectively changes the number of free parameters and allows for a more flexible description of smooth modulations in the PA across phase. Being degenerate, the parameter $a_0$ was fixed to 0. Fourier models of order $H = 2$ to $H=6$ were evaluated (corresponding to 4--12 free coefficients). A visual example of such models is provided on the right side of Figure~\ref{fig:lin_fou_models}.

To validate the modeling framework and assess the sensitivity to spurious features, extensive simulations were used. These included both high-statistics simulations with an artificially boosted flux (based on the model in \S\ref{sec:theo}, with $100\times$ the observed norm over 2 Ms), as well as simulations with realistic flux levels and an unpolarized source model. Both real and simulated data were phase-folded and extracted from the same source region. Comparisons between model families were performed on both the full phase interval and a restricted phase range (0.5–1), where features in the polarization signal might be more prominent. For each dataset, the linear and Fourier models were tested under similar sampling conditions to ensure fair comparison.

\begin{figure}
    \centering
    \includegraphics[width=0.45\linewidth]{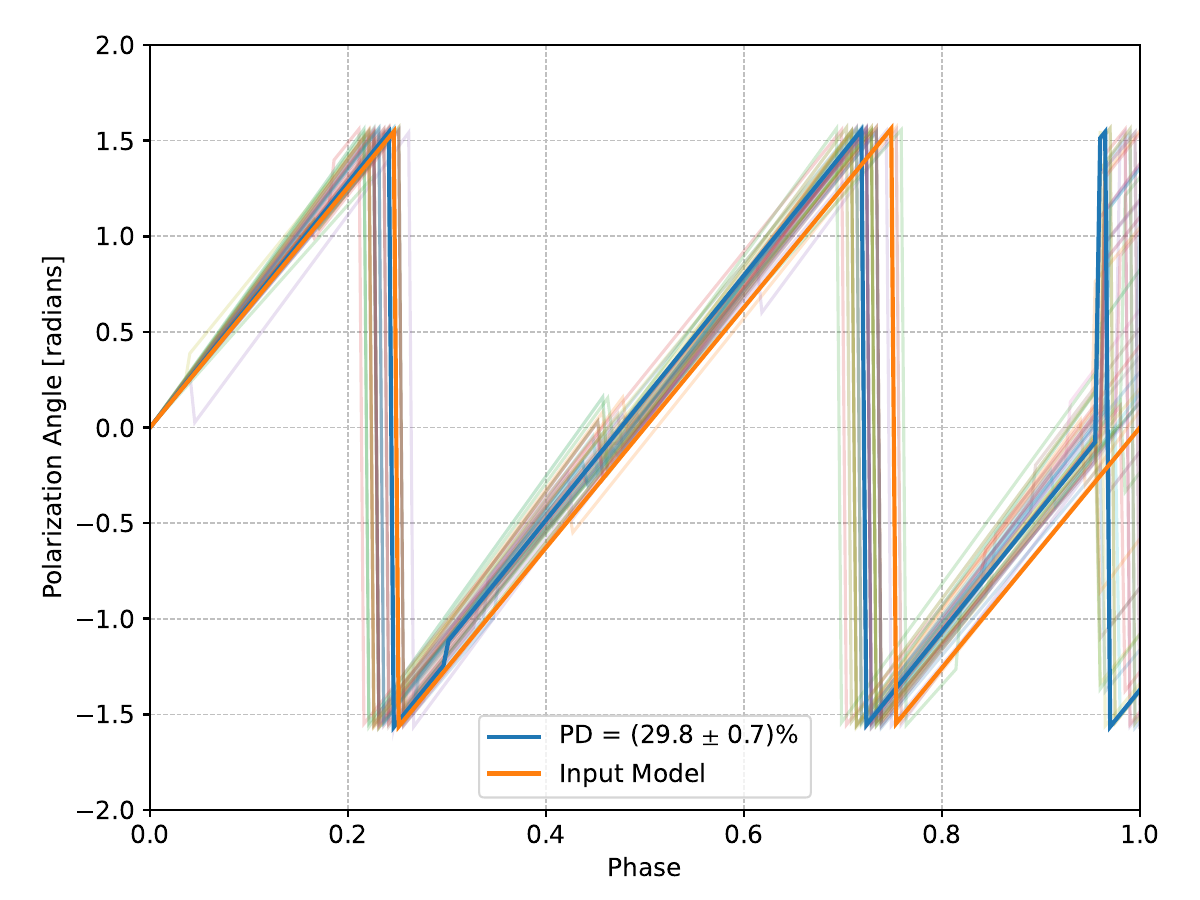}~~~~
    \includegraphics[width=0.45\linewidth]{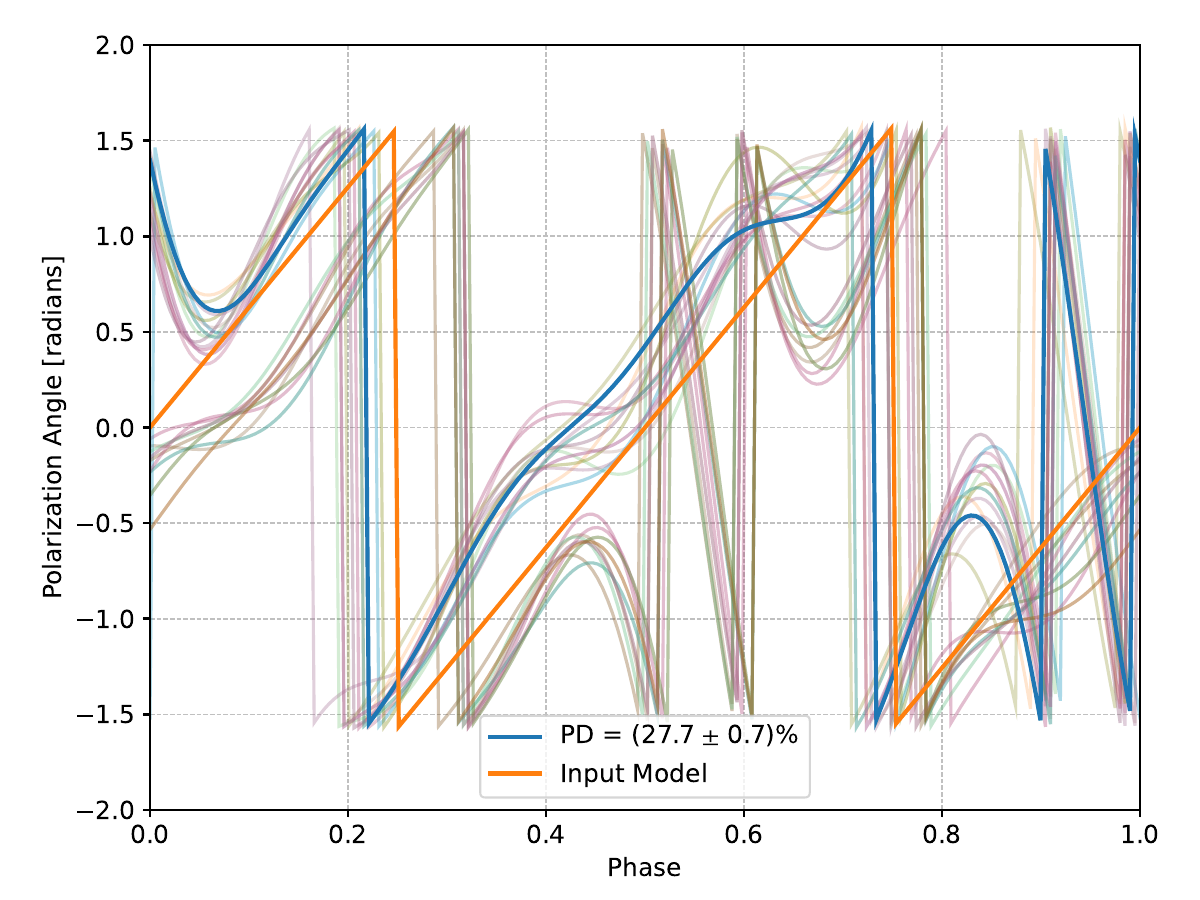}
    \includegraphics[width=0.45\linewidth]{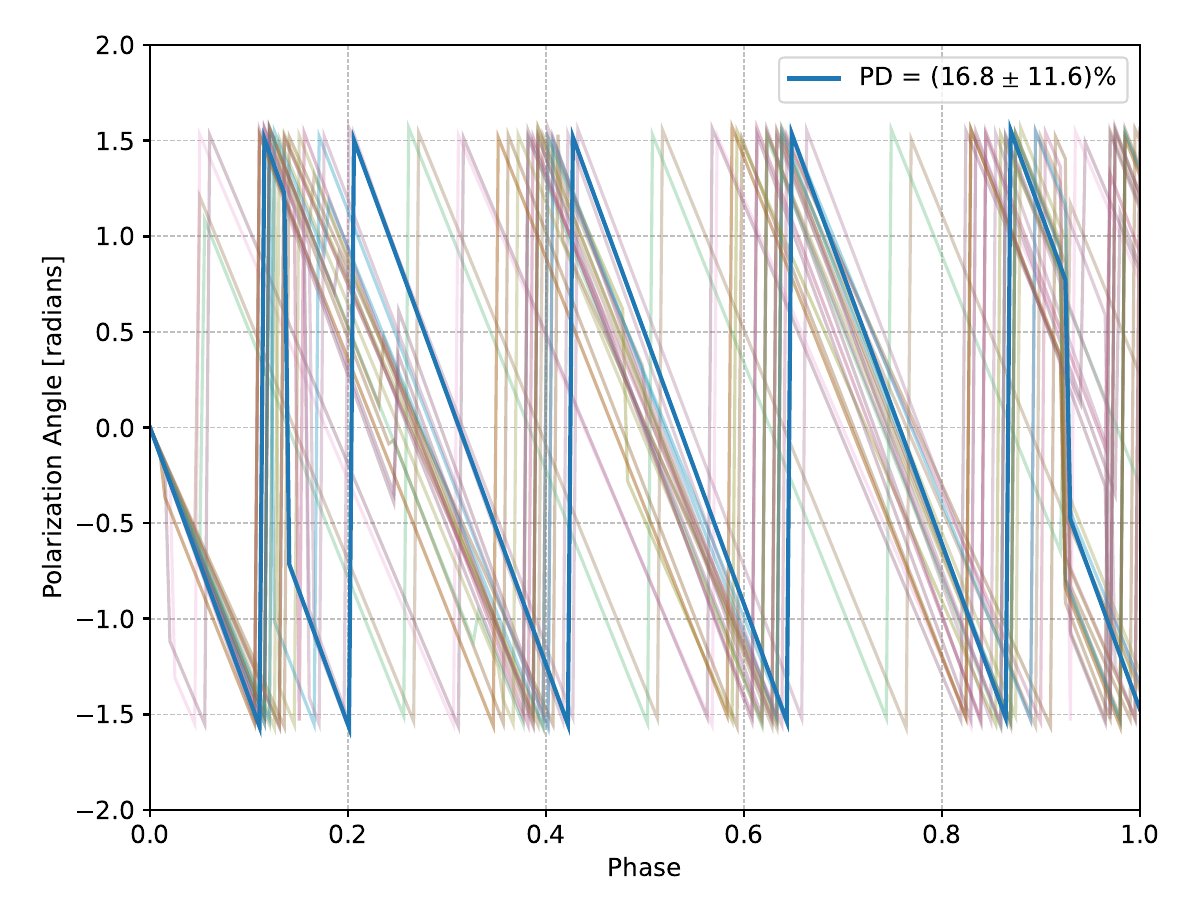}~~~~
    \includegraphics[width=0.45\linewidth]{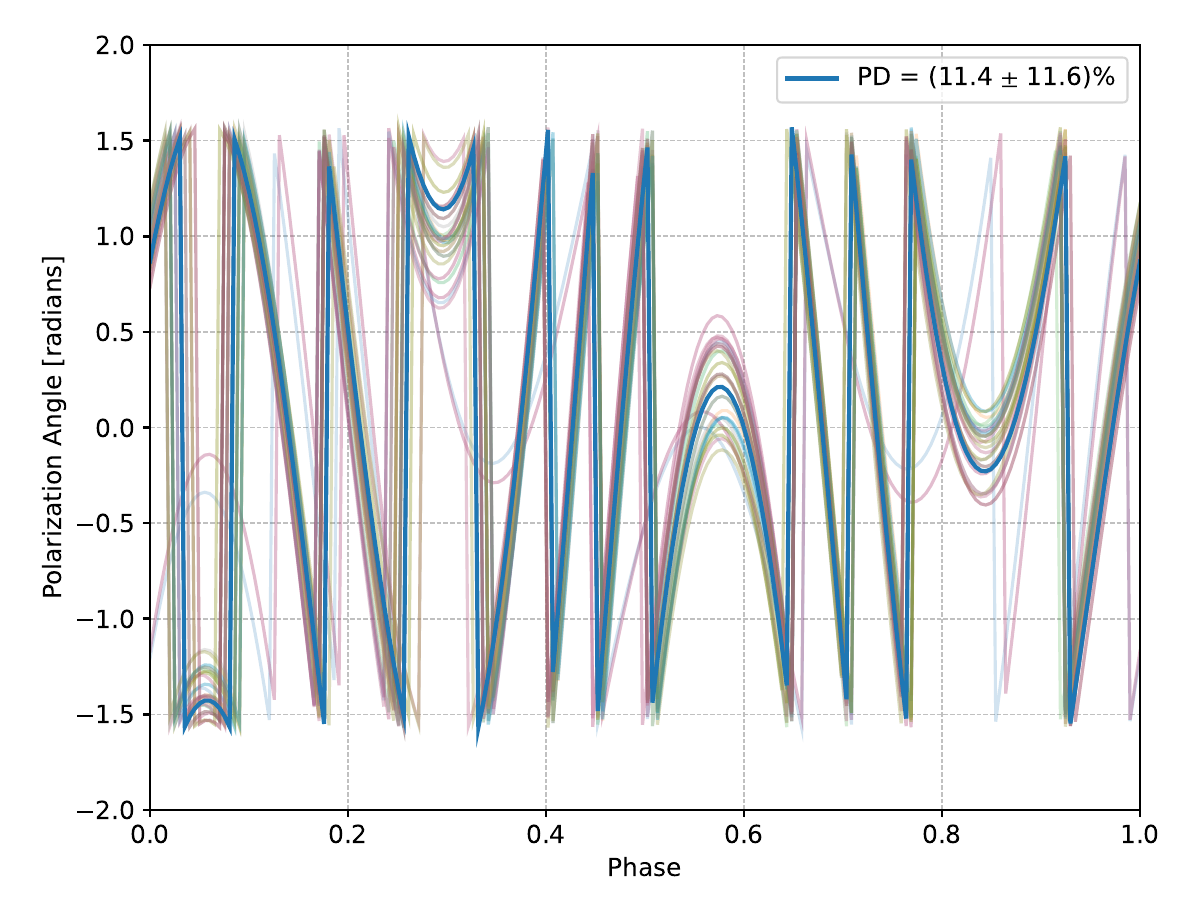}
    \includegraphics[width=0.45\linewidth]{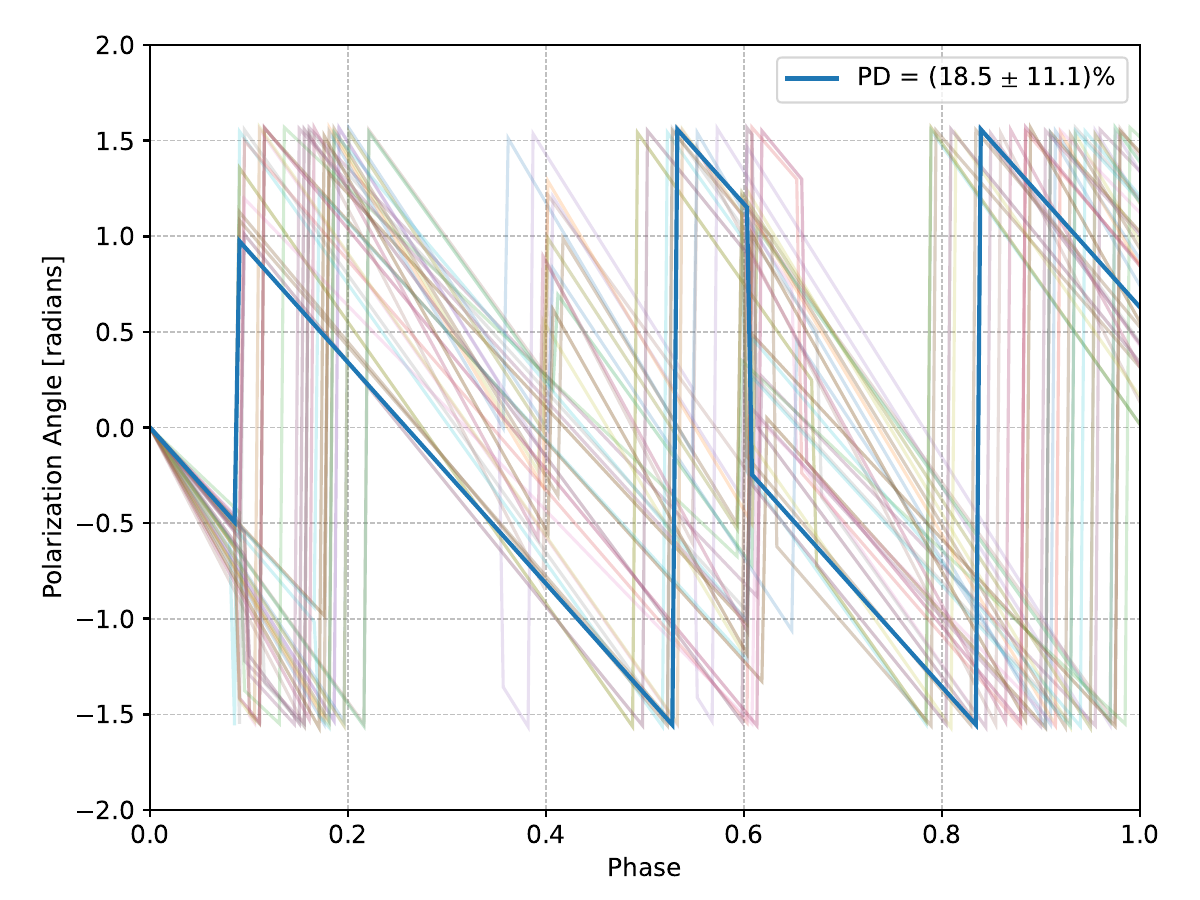}~~~~
    \includegraphics[width=0.45\linewidth]{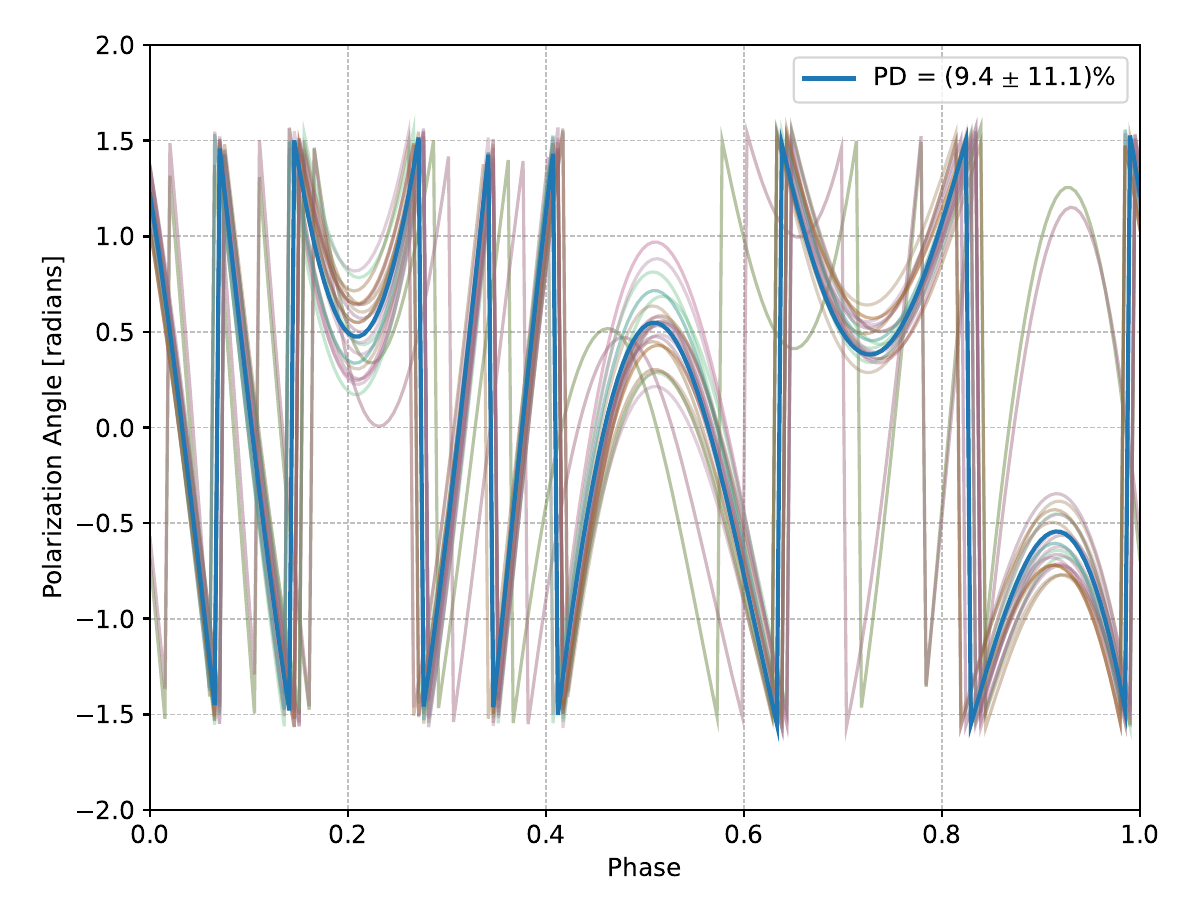}
    \caption{From top to bottom, the rows show the maximum-likelihood estimates of PA (wrapped within $\pm$90°) as a function of phase for (i) the polarized simulation with artificially boosted flux, (ii) the unpolarized simulation with realistic flux, and (iii) the real IXPE data. The left column displays results obtained with linear models, while the right column corresponds to Fourier models. The input PA model for the polarized simulation is shown in orange. In all six panels, the best-fit model is plotted in blue, and the other semi-transparent colored lines represent the next 50 most likely PA models.}
    \label{fig:sim_data_model_results}
\end{figure}

Figure~\ref{fig:sim_data_model_results} presents six panels showing the results of the maximum-likelihood analysis, obtained with linear models (left column) and Fourier models (right column). The rows correspond to the polarized simulation with boosted flux (top), the unpolarized simulation with realistic flux (middle), and the real IXPE data (bottom). For the boosted polarized simulation, the maximum-likelihood method correctly recovers the input polarization model, demonstrating its validity. This approach illustrates the potential for exploiting the phase dependence of PA to improve PD constraints. 
%However, the results from the real data remain ambiguous: some apparent features may simply arise from noise, as suggested by similar structures in the unpolarized simulations. 
The real IXPE data yield ambiguous results: while a negative slope close to –2$\pi$ appears to be favored in the linear models, a similar feature also arises in the unpolarized simulations, suggesting that it may be due to noise. Fourier models also show rapid oscillations in both real and unpolarized simulated data, consistent with statistical fluctuations. In all cases, the measured PD remains consistent with an unpolarized source.
Although no significant PA modulation is detected, this method may nevertheless prove valuable in future observations, either by IXPE or by possible new more sensitive missions (e.g. the Enhanced X-ray Timing and Polarimetry \citep[eXTP][]{2025SCPMA..6819502Z}), where increased statistics could help disentangle genuine physical signatures from statistical fluctuations.

\end{document}